\documentclass{aa}  

\usepackage{graphicx}
\usepackage{txfonts}
\usepackage[colorlinks=true,linkcolor=blue,citecolor=blue]{hyperref}

\begin{document} 

   \title{Formation-flying interferometry in geocentric orbits}

   \author{Takahiro Ito \inst{1}}

   \institute{Institute of Space and Astronautical Science, 3-1-1 Yoshinodai, Chuo-ward, Sagamihara, Kanagawa, 252-0222, Japan\\
              \email{ito.takahiro@jaxa.jp}
             }

 
  \abstract
   {Spacecraft formation flying serves as a method of astronomical instrumentation that enables the construction of large virtual structures in space. The formation-flying interferometry generally requires very high control accuracy, and extraterrestrial orbits are typically selected. To pave the way for comprehensive missions, proposals have been made for preliminary space missions, involving nano- or small satellites, to demonstrate formation-flying interferometry technologies, especially in low Earth orbits. From a theoretical perspective, however, it is unknown where and to what extent feasible regions for formation-flying interferometry should exist in geocentric orbits. }
   {This study aims to demonstrate the feasibility of formation-flying interferometry in geocentric orbits in which various perturbation sources exist. Geocentric orbits offer the advantage of economic accessibility and the availability of proven formation-flying technologies tailored for Earth orbits. Its feasibility depends on the existence of specific orbits that satisfy a small-disturbance environment with good observation conditions. }
   {Spacecraft motions in Earth orbits subjected to perturbations are analytically modeled based on celestial mechanics. The magnitudes of the accelerations required to counteract these perturbations are characterized by parameters such as the semimajor axis and the size of the formation. }
   {Small-perturbation regions tend to appear in higher-altitude and shorter-separation regions in geocentric orbits. Candidate orbits are identified in high Earth orbits for the triangular laser-interferometric gravitational-wave telescope, which is 100 km in size, and in medium Earth orbits for the linear astronomical interferometer, which is 0.5 km in size. A low Earth orbit with a separation of approximately 0.1 km may be suitable for experimental purposes.}
   {Geocentric orbits are potentially applicable for various types of formation-flying interferometry. }


   \maketitle
%

\section{Introduction} \label{Introduction}

Spacecraft formation flying represents a crucial method in astronomical instrumentation, with the potential to revolutionize modern astronomy by creating large virtual structures in space. Some space-based astronomical missions under conceptual study require rigid and continuous formation while maintaining a very high control accuracy of less than 1 $\mu$m and/or $\mu$rad, and interferometric techniques are typically used. The Deci-hertz Interferometer Gravitational Wave Observatory \citep[DECIGO;][]{kawamura2021current} is a laser-interferometric, space-based gravitational-wave observatory used to measure coalescing binary neutron stars, the acceleration of the universe, and stochastic gravitational waves predicted by inflation. Each cluster comprises three spacecraft that are 1,000 km apart and form an equilateral triangle, and four clusters are placed in a near-circular heliocentric orbit. Fabry-P\'{e}rot cavities are employed in DECIGO, each comprising two mirrors to measure the arm-length changes. The Large Interferometer For Exoplanets \citep[LIFE;][]{Quanz2022Large} is a space-based mid-infrared nulling interferometer that detects and characterizes the thermal emissions of exoplanets. LIFE is based on earlier mission concepts, such as Darwin \citep{Cockell2009darwin} and the Terrestrial Planet Finder Interferometer \citep{Lawson2005technology}, and its primary orbital selection is an orbit around the Sun-Earth Lagrange second (L2) point. The current design for LIFE assumes four collector spacecrafts and one beam-combiner spacecraft, and their formation size is controlled to remain between tens and hundreds of meters. The common features of DECIGO and LIFE are that precise control is necessary to maintain the optical path length or difference between satellites and that their candidate orbits are beyond Earth’s orbit. 

Flight-proven examples in formation flying have been shown in geocentric orbits, and coarse formation control on the order of submeters to meters was typically sufficient to satisfy mission objectives. The Japanese Engineering Test Satellite-VII \citep{Kawano2001result} demonstrated autonomous formation flying, rendezvous, and docking in the late 1990s in a circular low Earth orbit (LEO). In the 2000s and 2010s, autonomous formation-flying technology advanced for circular LEOs through leading space missions such as Prototype Research Instruments and Space Mission technology Advancement \citep[PRISMA;][]{Gill2007autonomous} and TerraSAR-X/TanDEM-X (TerraSAR-X add-on for Digital Elevation Measurement) \citep{Kahle2014formation} for technological demonstration and Earth observation. Their accuracy during formation flying reached the centimeter to meter level based on the navigation accuracy provided by the global navigation satellite system (GNSS). In the 2020s, Project for On-Board Autonomy-3 (PROBA-3) will demonstrate several formation activities, including coarse-to-fine control, autonomy, and safety management, to observe the solar coronagraph in a highly elliptical orbit. Its formation-flying precision is capable of submillimeter relative displacement and arcsecond-order pointing accuracy. This accuracy will be achieved using state-of-the-art metrology sensors and millinewton-class thrusters \citep{Penin2020proba3}. As seen in these examples, formation flight technologies are gradually maturing; however, a technological gap still exists in the development of future formation-flying interferometry. 

System-level, end-to-end, in-space verification for the pathfinder may be necessary to mature formation-flying interferometry before full-scale missions, as suggested in \citet{Monnier2019realistic}. However, the use of extraterrestrial orbits may be an inherent bottleneck for this purpose. The major benefits of using extraterrestrial orbits are the small-disturbance environment and good observation conditions. However, its major limitations are the high transportation cost to the desired orbit and that GNSS cannot be used for formation-flying autonomy, safety, and management. Such drawbacks may lead to formation-flying interferometry becoming a larger and longer-term project. 

An alternative is to use a geocentric orbit in which various perturbation sources exist. The expected advantages of using Earth orbits are the economic accessibility and availability of flight-proven technologies for formation flying autonomy, safety, and management with GNSS. Furthermore, state-of-the-art technologies are extending the use of GNSS from LEO to high Earth orbit \citep{Ashman2018gps}. However, few studies have applied Earth orbits to formation-flying interferometry. Representative and state-of-the-art research in this context conducted by \cite{hansen_ireland_2020} involved a linear astronomical interferometer in a near-circular LEO, and Earth’s $J_2$ gravity potential and atmospheric drag were considered to be major perturbations. The necessary $\Delta V$ (the time integral of the thrust acceleration) to maintain the strict formation requirements was at an acceptable level of a few tens of $\rm m\,s^{-1}$ per year, as calculated based on the numerical integration of the equations of motion. However, the numerical approach tends to rely on trial and error to determine feasible Earth-orbit regions. From a theoretical perspective, understanding where and to what extent feasible regions should exist is difficult. The characterization of orbital perturbations is as crucial for formation-flying interferometry as site characterization is for the deployment of ground-based telescopes. 

This study investigates the feasibility of formation-flying interferometry in geocentric orbits based on a celestial-mechanics approach. This study makes four major contributions to the literature. The first contribution is the analytical modeling of the perturbed orbital motion to attain precise formation flying. We identify four types of absolute and relative perturbations that need to be mitigated. The second contribution is to offer a propellant-efficient control strategy for correcting the absolute and/or relative perturbing accelerations. Secular and long-period perturbations of the eccentricity vector play key roles in determining the control strategy. The third contribution enables a complete and explicit calculation of major perturbing accelerations. Using the proposed method, the disturbance environments for different formations can be analytically evaluated in various Earth orbits. By integrating these three contributions, the perturbation magnitudes can be characterized by the semimajor axis and size of the formation, and this information can be used as a guideline for finding candidate orbits. The final contribution is the provision of detailed examples of orbital selection and control approaches for formation-flying interferometry. The proposed approach as well as methods for satisfying the observation conditions will be useful for future applications. 

This study considers only a near-circular Earth orbit because such is sufficient for various potential applications in formation-flying interferometry. In terms of terminology, this study uses the chief-deputy expression to describe formation-flying satellites. A chief does not necessarily represent the actual chief satellite, but rather the virtual origin of formation-flying satellites. By contrast, a deputy denotes a real spacecraft flying in proximity to the chief. These terms are used throughout the paper without any notation. Lastly, this study uses numerous symbols owing to the thorough analysis of perturbed motion. The major symbols, particularly those used in the theoretical development, are listed in Appendix \ref{Nomenclature}. 

The remainder of this paper is organized as follows. Section \ref{Unperturbed relative motion} presents a review of unperturbed relative motion for a circular orbit. Section \ref{Perturbed relative motion} presents the development of a perturbed relative-motion model for a near-circular orbit against arbitrary perturbations and explains the four types of resultant perturbing accelerations. Section \ref{Perturbation mitigation methods} first presents an analysis of the relationship between the perturbation periods and their contributions to perturbing acceleration magnitudes. Then, it presents analytical models of the perturbed orbital motions and perturbing accelerations as well as an effective approach for orbital selection and control. Section \ref{Results and discussion} presents numerical examples of the proposed approach to formation-flying interferometry. The concluding remarks are presented in Sect. \ref{Conclusions}.

\section{Unperturbed relative motion}\label{Unperturbed relative motion}
The relative equation of motion of the deputy with respect to the chief in the Earth-centered inertial (ECI) frame is
\begin{equation}
    \ddot{\boldsymbol{r}}^I= - \mu_e\left(\frac{\boldsymbol{R}_d^I}{R_d^3}-\frac{\boldsymbol{R}_c^I}{R_c^3}\right) + {\boldsymbol{f}^I_p} + (\boldsymbol{u}_d^I-\boldsymbol{u}^I_c),\label{relative_eom_eci}
\end{equation}
where $\mu_e$ is the Earth gravitational constant, $\boldsymbol{r}^I$ is the relative position vector of the deputy ($\boldsymbol{r}^I=\boldsymbol{R}_d^I-\boldsymbol{R}_c^I$), $\boldsymbol{R}_c^I$ and $\boldsymbol{R}_d^I$ are the position vectors of the chief and deputy, respectively, {$\boldsymbol{f}^I_p$} is the disturbance acceleration vector of the deputy relative to the chief (${\boldsymbol{f}^I_p}=\boldsymbol{F}_d^I-\boldsymbol{F}^I_c$), $\boldsymbol{F}^I_c$ and $\boldsymbol{F}_d^I$ are the disturbance acceleration vectors for the chief and deputy, respectively, $\boldsymbol{u}^I_c$ and $\boldsymbol{u}_d^I$ are the control acceleration vectors applied to the chief and deputy, respectively, $R_c=|| \boldsymbol{R}_c^I ||$, and $R_d=|| \boldsymbol{R}_d^I ||$.  The superscript $I$ indicates the position, velocity, or acceleration in the ECI frame.

By assuming the unperturbed motion of both the chief and deputy (${\boldsymbol{f}^I_p}=\boldsymbol{0}$) and the uncontrolled motion of the chief ($\boldsymbol{u}^I_c=\boldsymbol{0}$), Eq. (\ref{relative_eom_eci}) in the ECI frame is transformed into the local-vertical-local-horizontal (LVLH) frame: 
\begin{eqnarray}
\left\{
\begin{aligned}
 \ddot{x} - 2 \dot{\theta}\dot{y} - \ddot{\theta} y - \dot{\theta}^2 x &= - \frac{\mu_e (R_c + x)}{R_d^3} + \cfrac{\mu_e}{R_c^2}+u_{x}, \\
        \ddot{y} + 2 \dot{\theta} \dot{x} + \ddot{\theta} x - \dot{\theta}^2 y &= - \frac{\mu_e y}{R_d^3}+u_{y}, \\
        \ddot{z} &= - \cfrac{\mu_e z}{R_d^3}+u_{z}, 
\end{aligned}\label{CW_nonlinear}
\right.
\end{eqnarray}
where $\boldsymbol{r}=[x,y,z]^T$ denotes the position vector of the deputy relative to the chief in the LVLH frame, $\theta$ is the argument of latitude of the chief, and $\boldsymbol{u}_d=[u_{x},u_{y},u_{z}]^T$ is the control acceleration vector applied to the deputy in the LVLH frame.  The position, velocity, or acceleration in the LVLH frame are written without superscript notations, unlike those in the ECI frame. The LVLH frame is defined such that the $x$ direction is directed radially outward from the chief, the $z$ direction is normal to the orbital plane and positive in the direction of the (instantaneous) angular momentum, and the $y$ direction in the right-hand frame completes the setup. Assuming $r \ll R_c$ ($r=|| \boldsymbol{r} ||$), the first term on the right-hand side of Eq. (\ref{CW_nonlinear}) can be linearized as follows:
\begin{eqnarray}
\left\{
\begin{aligned}
        - \frac{\mu_e (R_c + x)}{R_d^3} &\simeq \frac{\mu_e(2x-R_c)}{R_c^3},\\
        - \frac{\mu_e y}{R_d^3} &\simeq - \frac{\mu_e y}{R_c^3},\\
        - \frac{\mu_e z}{R_d^3} &\simeq - \frac{\mu_e z}{R_c^3}. 
\end{aligned}\label{HCW_linearization} 
\right.
\end{eqnarray}
For a circular orbit, $\dot{\theta}=n$, $\ddot{\theta}=0$, and $R_c=a$, where $n$ and $a$ are the mean motion and semimajor axis of the chief, respectively. In addition, using Eqs. (\ref{CW_nonlinear}) and (\ref{HCW_linearization}) and the relationship $n=\sqrt{\mu_e/a^3}$, the unperturbed and linear relative dynamics model for a circular orbit is obtained as follows:
\begin{equation}
\ddot{\boldsymbol{r}}={A}_{u1}\boldsymbol{r}+{A}_{u2}\dot{\boldsymbol{r}}+\boldsymbol{u}_d, \label{system_equation}
\end{equation}
where ${A}_{u1}$ and ${A}_{u2}$ are matrices defined as follows:
\begin{equation}
{A}_{u1}=\begin{bmatrix}
3n^2&0&0\\
0&0&0\\
0&0&-n^2\\
\end{bmatrix}, \quad
{A}_{u2}=
\begin{bmatrix}
0&2n&0\\
-2n&0&0\\
0&0&0\\
\end{bmatrix}.
\label{A_CW}
\end{equation}
Equation (\ref{system_equation}) is known as the Clohessy--Wiltshire (CW) equation \citep{Clohessy1960terminal}. 

A possible approach for realizing formation flying under the orbital dynamics in Eq. (\ref{system_equation}) is to cancel the relative accelerations ${A}_{u1}\boldsymbol{r}$ and ${A}_{u2}\dot{\boldsymbol{r}}$ with the control accelerations. However, the control acceleration requirement tends to become large, which is inefficient and unrealistic for formation-flying applications. Therefore, the reference formation orbit is typically selected using the natural solution of Eq. (\ref{system_equation}) when $\boldsymbol{u}_d=\boldsymbol{0}$. The bounded CW solution (the solution without an along-track drift) is 
\begin{equation}
\boldsymbol{r}_r=
\begin{bmatrix}
\rho_x \sin{(\theta+\alpha_x)}\\
\rho_y + 2\rho_x \cos{(\theta+\alpha_x)}\\
\rho_z \sin{(\theta+\alpha_z)}\\
\end{bmatrix},\label{CW_bounded_solution}
\end{equation}
where $\rho_x$, $\rho_y$, $\rho_z$, $\alpha_x$, and $\alpha_z$ denote the design parameters of the relative orbit \citep{Clohessy1960terminal}.  Clearly, $\theta$ is a linear function of time because $\dot{\theta}=n$ for the unperturbed circular orbit; therefore, $\boldsymbol{r}_r$ is also a function of time. 

\section{Perturbed relative motion}\label{Perturbed relative motion}
In contrast to Keplerian dynamics, oscillating and drifting orbital motion can arise in the presence of orbital perturbations. The reference formation described in the previous section was obtained using unperturbed, circular, and linearized assumptions. Therefore, elliptical and nonlinear effects may disturb the actual formation. This section analyzes how orbital perturbations and un-modeled dynamics can affect the rigid and continuous formation control.

\begin{table*}
\caption{\label{table1} Categories of perturbing accelerations that can be mitigated by the control. }
\centering
\begin{tabular}{ll}
\hline \hline 
Category & Description\\ \hline
$\boldsymbol{F}_c$ & Absolute physical perturbation. \\ 
$\boldsymbol{f}_p$ & Relative physical perturbation. \\ 
$\boldsymbol{f}_o$ & Relative fictitious perturbation owing to the perturbed orbital motion of the chief. \\ 
$\boldsymbol{f}_f$ & Relative fictitious perturbation owing to the perturbed motion of the reference formation.  \\ \hline
\end{tabular}
\end{table*}

The perturbed relative equation of motion of the deputy in the LVLH frame is expressed as follows \citep{Izzo2003new}:
\begin{equation}
\ddot{\boldsymbol{r}}=-2\boldsymbol{\omega} \times \dot{\boldsymbol{r}} - \boldsymbol{\omega} \times (\boldsymbol{\omega} \times \boldsymbol{r}) - \dot{\boldsymbol{\omega}} \times \boldsymbol{r} + {\boldsymbol{f}_{g2B} + \boldsymbol{f}_p}+\boldsymbol{u}_d - \boldsymbol{u}_c, \label{Perturbed_Equation}
\end{equation}
where $\boldsymbol{\omega}=[\omega_x,\omega_y,\omega_z]^T$ is the orbital angular velocity vector of the chief in the LVLH frame, {$\boldsymbol{f}_{g2B}$} is the linearized gravity gradient acceleration owing to the two-body gravitational field (${\boldsymbol{f}_{g2B}}=-(\mu_e/R_c^3)[-2x,y,z]^T$), {$\boldsymbol{f}_p$} is the sum of the relative and physical perturbing accelerations applied to the deputy, and $\boldsymbol{u}_c$ and $\boldsymbol{u}_d$ are the control acceleration vectors applied to the chief and deputy in the LVLH frame, respectively. The elements of $\boldsymbol{\omega}$ are \citep{Alfriend2010spacecraft}
\begin{eqnarray}
\left\{
\begin{aligned}
\omega_x&=\dot{\Omega}\sin{I}\sin{\theta}+\dot{I}\cos{\theta}, \\
\omega_y&=\dot{\Omega}\sin{I}\cos{\theta}-\dot{I}\sin{\theta}=0, \\
\omega_z&=\dot{\Omega}\cos{I}+\dot{\theta}, 
\end{aligned}\label{angvel}
\right.
\end{eqnarray}
where $\Omega$ and $I$ are the right ascension of the ascending node (RAAN) and inclination of the chief, respectively. We can verify $\omega_y=0$ by substituting $\dot{I}$ and $\dot{\Omega}$ from Gauss's variational equations into Eq. (\ref{angvel}) \citep{Alfriend2010spacecraft}. 

Next, Eq. (\ref{Perturbed_Equation}) is transformed into
\begin{equation}
\ddot{\boldsymbol{r}}={A}_{t1}\boldsymbol{r}+{A}_{t2}\dot{\boldsymbol{r}}
+{\boldsymbol{f}_p+\boldsymbol{f}_o}+\boldsymbol{u}_d - \boldsymbol{u}_c. \label{Perturbed_Equation2}
\end{equation}
We note that ${A}_{t1}$ and ${A}_{t2}$ are $(3\times3)$ matrices, where the elements of ${A}_{u1}$ and ${A}_{u2}$ in Eq. (\ref{A_CW}) are replaced by $n \xrightarrow{}\omega_t$, where $\omega_t$ is the user-defined target angular velocity. In Eq. (\ref{Perturbed_Equation2}), {$\boldsymbol{f}_o$} takes the following form:
\begin{eqnarray}
{\boldsymbol{f}_o}=({A}_{p1}-{A}_{t1})\boldsymbol{r}+({A}_{p2}-{A}_{t2})\dot{\boldsymbol{r}},\label{F_o}
\end{eqnarray}
where ${A}_{p1}$ and ${A}_{p2}$ are $(3\times3)$ matrices corresponding to the first four terms in Eq. (\ref{Perturbed_Equation}). It should be noted that {$\boldsymbol{f}_o$} is the relative fictitious acceleration owing to the perturbed orbital motion of the chief, and it remains zero if the chief orbit is unperturbed and circular. Equation (\ref{F_o}) suggests that the selection of the $\omega_t$ profile is important for minimizing {$\boldsymbol{f}_o$}. Choosing $\omega_t$ close to $\omega_z$ is one of the best solutions, and Eq. (\ref{F_o}) can be approximated further. Hereafter, $\omega_t$ is selected as $\omega_z$ throughout this study. First, under the near-circular-orbit assumption, the argument of latitude can be approximated as follows: 
\begin{equation}
    \theta = (\phi+2e_{x}\sin{\phi}-2e_{y}\cos{\phi}),\label{theta_app_e}
\end{equation}
where $\phi$ is the mean argument of latitude defined as $\phi=(l+\omega)$, $\omega$ is the argument of perigee, $l$ is the mean anomaly, $e_{x}$ and $e_{y}$ are the elements of the eccentricity vector defined as $[e_{x},e_{y}]^T=[e\cos{\omega}, e\sin{\omega}]^T$, and $e$ is the eccentricity of the chief. Then, let the time derivative of $\phi$ be expressed as $\dot{\phi} = (n+\Delta \dot{\phi})$. Moreover, the time derivatives $\Delta \dot{\phi}$, $\dot{I}$, and $\dot{\Omega}$ are assumed sufficiently small with respect to $n$. By neglecting the second- and higher-order terms of the small variables, 
$({A}_{p1}-{A}_{t1})$ and $({A}_{p2}-{A}_{t2})$ in Eq. (\ref{F_o}) can be expressed as follows:
\begin{eqnarray}
\begin{aligned}
({A}_{p1}-{A}_{t1})&=
    \begin{bmatrix}
        a^{(11)}_{A1} & a^{(12)}_{A1} & a^{(13)}_{A1}\\
        -a^{(12)}_{A1} & a^{(22)}_{A1} & a^{(23)}_{A1}\\
         a^{(13)}_{A1} &-a^{(23)}_{A1} & a^{(33)}_{A1}
    \end{bmatrix}, \\
    ({A}_{p2}-{A}_{t2})&=
    \begin{bmatrix}
        0& 0 & 0\\
        0& 0 & a^{(23)}_{A2}\\
        0&-a^{(23)}_{A2} & 0\\
    \end{bmatrix},
\end{aligned} \label{dAp_mat}
\end{eqnarray}
where
\begin{eqnarray}
\begin{aligned}
a^{(11)}_{A1}=&{ -4n(\dot{\Omega}\cos{I}+\Delta \dot{\phi}+2\dot{e}_{x}\sin{\phi}-2\dot{e}_{y}\cos{\phi})}\\
&{-2n^2(e_{x}\cos{\phi}+e_{y}\sin{\phi}),}\\
a^{(12)}_{A1}=&\ddot{\Omega}\cos{I}{+\ddot{\phi}+2[(\ddot{e}_{x}+2n\dot{e}_{y}-n^2e_{x})\sin{\phi}}\\
&{-(\ddot{e}_{y}-2n \dot{e}_{x}-n^2e_{y})\cos{\phi}], }\\
a^{(13)}_{A1}=&-n (\dot{\Omega}\sin{I}\sin{\phi}+\dot{I}\cos{\phi}), \quad
a^{(22)}_{A1}=a^{(33)}_{A1}=-\cfrac{1}{2}a^{(11)}_{A1}, \\ 
a^{(23)}_{A1}=&(\ddot{\Omega}\sin{I}-n\dot{I})\sin{\phi}+(\ddot{I}+n\dot{\Omega}\sin{I})\cos{\phi}, \\
a^{(23)}_{A2}=&2(\dot{\Omega}\sin{I}\sin{\phi}+\dot{I}\cos{\phi}).
\end{aligned}\label{dAp_element}
\end{eqnarray}
The following relationships are used to obtain Eqs. (\ref{dAp_mat}) and (\ref{dAp_element}): 
\begin{eqnarray}
\begin{aligned}
R_c=&a(1-e^2)(1+e_{x}\cos{\theta}+e_{y}\sin{\theta})^{-1},\\
R_c^{-3} \simeq& a^{-3}(1+3e_{x}\cos{\phi}+3e_{y}\sin{\phi}). 
\end{aligned}
\end{eqnarray}

Then, the desired reference formation is selected as the natural solution of Eq. (\ref{CW_bounded_solution}): 
\begin{equation}
\boldsymbol{r}_r=
\begin{bmatrix}
\rho_x \sin{\Theta_x}\\
\rho_y + 2\rho_x \cos{\Theta_x}\\
\rho_z \sin{\Theta_z}\\
\end{bmatrix},
\end{equation}
where $\Theta_x$ and $\Theta_z$ are the new variables satisfying $\Theta_x=(\theta+\alpha_x)$, $\Theta_z=(\theta+\alpha_z)$, $\dot{\Theta}_x=\dot{\Theta}_z=\omega_t$, and $\ddot{\Theta}_x=\ddot{\Theta}_z=\dot{\omega}_t$, and the parameters of ($\rho_x$, $\rho_y$, $\rho_z$) can be time-variant. 
The relationships between $\boldsymbol{r}_r$, $\dot{\boldsymbol{r}}_r$, and $\ddot{\boldsymbol{r}}_r$ can be written as follows:
\begin{equation}
\ddot{\boldsymbol{r}}_r+{\boldsymbol{f}_f}={A}_{t1}\boldsymbol{r}_r+{A}_{t2}{\dot{\boldsymbol{r}}}_r,\label{F_f}
\end{equation}
where
\begin{eqnarray}
\begin{aligned}
    {\boldsymbol{f}_{f}}=&
    {A}_{t2}\begin{bmatrix}
    \dot{\rho}_x\sin{\Theta_x}\\
    \dot{\rho}_y+2\dot{\rho}_x \cos{\Theta_x}\\
    \dot{\rho}_z\sin{\Theta_z}
    \end{bmatrix}
    -\dot{\omega}_t
    \begin{bmatrix}
    \rho_x\cos{\Theta_x}\\
    -2\rho_x \sin{\Theta_x}\\
    \rho_z\cos{\Theta_z}\\
    \end{bmatrix} \\
    &-2\omega_t
    \begin{bmatrix}
    \dot{\rho}_x\cos{\Theta_x}\\
    - 2\dot{\rho}_x \sin{\Theta_x}\\
    \dot{\rho}_z\cos{\Theta_z}
    \end{bmatrix}
    -\begin{bmatrix}
    \ddot{\rho}_x\sin{\Theta_x}\\
    \ddot{\rho}_y+2\ddot{\rho}_x \cos{\Theta_x}\\
    \ddot{\rho}_z\sin{\Theta_z}
    \end{bmatrix}.\label{F_f2}
\end{aligned}
\end{eqnarray}
We note that {$\boldsymbol{f}_f$} is the relative fictitious acceleration owing to the perturbed motion of the reference formation.

Finally, by subtracting Eq. (\ref{F_f}) from Eq. (\ref{Perturbed_Equation2}), we obtain the following:
\begin{equation}
\ddot{\boldsymbol{\epsilon}}={A}_{t1}\boldsymbol{\epsilon}+{A}_{t2}{\dot{\boldsymbol{\epsilon}}}
+{\boldsymbol{f}_p+\boldsymbol{f}_o+\boldsymbol{f}_f}+\boldsymbol{u}_d - \boldsymbol{u}_c,\label{Perturbed_Equation3}
\end{equation}
where $\boldsymbol{\epsilon}=(\boldsymbol{r}-\boldsymbol{r}_r)$ is the relative positional error with respect to the reference formation. By setting the control command of the deputy to
\begin{equation}
\boldsymbol{u}_d = \boldsymbol{u}_c+\boldsymbol{u}_r,\quad \boldsymbol{u}_r = -({\boldsymbol{f}_p+\boldsymbol{f}_o+\boldsymbol{f}_f}),\label{u}
\end{equation}
where $\boldsymbol{u}_r$ is the control acceleration to compensate for all relative disturbances, and by assuming that the current relative position and velocity are on the desired formation ($\boldsymbol{r}=\boldsymbol{r}_r$, $\dot{\boldsymbol{r}}=\dot{\boldsymbol{r}}_r$), the relative acceleration error $\ddot{\boldsymbol{\epsilon}}$ becomes zero; thus, tracking on the reference formation is expected for the deputy.

 Table \ref{table1} summarizes the four categories of perturbing accelerations that can be mitigated by the control. The simplest way to nullify the perturbation effects is to compensate for all absolute and relative perturbing accelerations (i.e., $\boldsymbol{F}_c$, which is the expression of $\boldsymbol{F}^I_c$ in the LVLH frame, $\boldsymbol{f}_p$, $\boldsymbol{f}_o$, and $\boldsymbol{f}_f$) using the control $\boldsymbol{u}_d$. However, compensation for absolute perturbing accelerations typically leads to a significant waste of propellant. In such cases, another possible approach is to maintain most of the absolute perturbing accelerations, except for the small ones, and compensate for all the relative perturbations with respect to the chief. The feasibility of such a control strategy depends on whether the control acceleration $\boldsymbol{u}_d$, and eventually $\boldsymbol{F}_c$, {$\boldsymbol{f}_p$, $\boldsymbol{f}_o$, and $\boldsymbol{f}_f$} are sufficiently small for the reference formation. Determining the differential disturbances of the deputy with respect to the chief is necessary to evaluate {$\boldsymbol{f}_p$}. Moreover, determining the perturbed orbital motions of ($e_{x}$, $e_{y}$) as well as the first and second time derivatives of ($e_{x}$, $e_{y}$, $I$, $\Omega$, $\phi$) is necessary to evaluate {$\boldsymbol{f}_o$}. The $\omega_t(=\omega_z)$ profile affects {$\boldsymbol{f}_o$} and {$\boldsymbol{f}_f$}, and understanding the perturbed motions of $\omega_z$ and $\dot{\omega}_z$, which are functions of ($e_{x}$, $e_{y}$, $I$, $\Omega$, $\phi$), is essential. Furthermore, the control-effort trade-off between compensating for or maintaining $\boldsymbol{F}_c$ remains unknown. Such concerns motivate the development of analytical models of perturbed orbital motions and perturbing accelerations, as described in the next section. 

\section{Perturbation mitigation methods}\label{Perturbation mitigation methods}

\subsection{Perturbation periods and acceleration magnitudes}\label{Perturbation periods and acceleration magnitudes}
The physical perturbing accelerations (i.e., $\boldsymbol{F}_c$ and $\boldsymbol{f}_p$) are determined simply from the positions and velocities of the chief and deputy, in addition to spacecraft and environmental parameters at a certain moment. In contrast, to calculate the fictitious perturbing accelerations (i.e., $\boldsymbol{f}_o$ and $\boldsymbol{f}_f$), it is necessary to understand the perturbed orbital motion of the chief, as presented in the previous section. In addition, the control accelerations to compensate for the absolute physical perturbations ($\boldsymbol{F}_c$) can be divided into the secular, long-period, and short-period parts through Gauss's variational equations. Whereas compensation for all parts of $\boldsymbol{F}_c$ could lead to a significant waste of propellant, compensation for small parts of $\boldsymbol{F}_c$ may be comparable to the relative fictitious perturbations. Therefore, before developing the perturbation models for each perturbing source, we show an analysis to understand how the perturbed orbital motion of the chief can affect the control accelerations to compensate for the relative fictitious perturbations, and how the relative fictitious perturbations are compared to the absolute physical ones with respect to the perturbation periods.

\begin{table*}
\caption{\label{table2} Orders of the control acceleration magnitudes to compensate for the relative fictitious perturbations ($\boldsymbol{f}_o$ and $\boldsymbol{f}_f$) for the secular, long-period, and short-period terms, in comparison to those of the absolute physical perturbations ($\boldsymbol{F}_c$).} 
\centering
\begin{tabular}{llll}
\hline \hline 
Terms & Secular & Long-period & Short-period \\ \hline
$\ddot{(\cdot)}r_r$ & 0 & $c^{(lp)} \times (r_r/a)(n_l/n)$ & $c^{(sp)} \times (r_r/a)$ \\ 
$\dot{(\cdot)} n r_r$ & $c^{(s)} \times (r_r/a)$ & $c^{(lp)} \times (r_r/a)$ & $c^{(sp)} \times (r_r/a)$ \\ 
${(\cdot)} n^2 r_r$ & $c^{(s)} \times (r_r/a)(n t)$ & $c^{(lp)} \times (r_r/a)(n_l/n)^{-1}$ & $c^{(sp)} \times (r_r/a)$ \\ \hline
\end{tabular}
\tablefoot{In the first column, $(\cdot)$ denotes the osculating orbital elements comprising the secular, long-period, and short-period terms as expressed in Eq. (\ref{OE_osc}). The osculating orbital elements of ($e_{x}$, $e_{y}$) are substituted to $(\cdot) n^2 r_r$, and those of ($e_{x}$, $e_{y}$, $I$, $\Omega$, $\phi$) are substituted to $\dot{(\cdot)} n r_r$ and $\ddot{(\cdot)} r_r$ for the analysis. In the second to fourth columns, $c^{(s)}$, $c^{(lp)}$, and $c^{(sp)}$ indicate the orders of the control acceleration magnitudes to compensate for the absolute physical perturbations for the secular, long-period, and short-period terms, respectively. They are defined as $c^{(s)}=\dot{\bar{(\cdot)}} n a$, $c^{(lp)}=\delta{(\cdot)^{(lp)}} n_{l} n a$, and $c^{(sp)}=\delta{(\cdot)^{(sp)}} n^2 a$, respectively. }
\end{table*}

Equations (\ref{F_o}), (\ref{dAp_mat}), and (\ref{dAp_element}) suggest that the osculating orbital motion produces the relative fictitious acceleration of $\boldsymbol{f}_o$, whose magnitude comprises the following terms: $(\cdot) n^2 r_r$ for ($e_{x}$, $e_{y}$), and $\dot{(\cdot)} n r_r$ and $\ddot{(\cdot)} r_r$ for ($e_{x}$, $e_{y}$, $I$, $\Omega$, $\phi$), where $(\cdot)$ expresses the osculating orbital element and $||\boldsymbol{r}_r||=r_r$. From Gauss's variational equations, the control accelerations to compensate for the absolute perturbations comprise the following terms: $\dot{(\cdot)} n a$ for ($e_{x}$, $e_{y}$, $I$, $\Omega$, $\phi$). At this point, the osculating orbital elements are modeled as follows:
\begin{equation}
    (\cdot)=(\cdot)^{(s)}+\delta (\cdot)^{(lp)}+\delta (\cdot)^{(sp)},\quad 
    (\cdot)^{(s)}={(\cdot)}^{(s)}_0+\dot{\bar{(\cdot)}}t, \label{OE_osc}
\end{equation}
where $(\cdot)^{(s)}$ is the secular term that is a linear function of time, $\delta (\cdot)^{(lp)}$ is the long-period term for which the angular frequency has an order of $n_l(\ll n)$, $\delta (\cdot)^{(sp)}$ is the short-period term for which the angular frequency has an order of $n$, ${(\cdot)}^{(s)}_0$ is the initial value of the secular term, and $\dot{\bar{(\cdot)}}$ is the drift rate of the secular term. By substituting Eq. (\ref{OE_osc}) into relevant terms, the effects of the secular, long-period, and short-period terms can be analyzed. The same discussion holds for the relative fictitious acceleration of $\boldsymbol{f}_f$, provided that $\rho_x$, $\rho_y$, and $\rho_z$ are derived from the natural solution in Eq. (\ref{CW_bounded_solution}).

Table \ref{table2} summarizes the orders of the control acceleration magnitudes required to compensate for the relative fictitious perturbations ($\boldsymbol{f}_o$ and $\boldsymbol{f}_f$) for the secular, long-period, and short-period terms in comparison with those of the absolute physical perturbations ($\boldsymbol{F}_c$). Table \ref{table2} suggests that most terms to compensate for the relative fictitious perturbations include the common terms of the absolute physical ones multiplied by $(r_r/a)$ and/or $(n_l/n)$, both of which are sufficiently smaller than 1. Therefore, the best approach is to compensate for the relative fictitious perturbations to gain less control acceleration. However, two exceptional cases appear in the secular and long-period parts of the term with ${(\cdot)} n^2 r_r$. They have common terms of the absolute physical perturbations multiplied by not only $(r_r/a)$ ($\ll 1$), but also $(n t)$, which can be considerably larger than 1 when $t$ is large, or $(n_l/n)^{-1}$, which is significantly larger than 1. This result implies that the secular and long-period motions of the eccentricity vector ($e_{x}$, $e_{y}$) must be compared to determine whether to compensate for the absolute or relative perturbations to attain more efficient control. This insight is used to determine the orbit and formation control approach in Sect. \ref{Control approach}. 

\subsection{Analytical perturbation models}\label{Analytical perturbation models}
The perturbation sources considered in this study are CW nonlinearity, Earth $J_2$ gravity potential, Earth $J_3$ gravity potential, lunisolar gravity, atmospheric drag, and  SRP. Their analytical models are explained in the following. 

\subsubsection{CW nonlinearity}
The relative acceleration owing to CW nonlinearity is produced by nonlinear terms that are neglected in the CW equation (Eq. (\ref{system_equation})). This can be computed by subtracting the first term on the right-hand side of Eq. (\ref{HCW_linearization}) from the left-hand side. The CW nonlinearity can contribute to a part of $\boldsymbol{f}_p$.

\subsubsection{Earth $J_2$ gravity potential}
The $J_2$ gravity potential of Earth’s gravity is known to be one of the major perturbation sources for satellites, particularly in low-to-medium Earth orbit. By considering only zonal effects, the $J_2$-perturbed absolute acceleration vector acting on the chief ($\boldsymbol{F}_{J_2}$, as a part of $\boldsymbol{F}_c$) is given by \citep{Alfriend2010spacecraft}
\begin{equation}
    \boldsymbol{F}_{J_2}=-\cfrac{3J_2\mu_eR_e^2}{2R^4_c}
    \begin{bmatrix}
        1-3\sin^2{I}\sin^2{\theta}\\
        \sin^2{I}\sin{2\theta}\\
        \sin{2I}\sin{\theta}\\
    \end{bmatrix},\label{fc_j2}
\end{equation}
where $R_e$ is the mean equatorial radius of Earth. The linearized $J_2$ differential acceleration vector ($\boldsymbol{f}_{J_2}$, as a part of $\boldsymbol{f}_p$) has the following form \citep{Izzo2003new}:
\begin{equation}
    {\boldsymbol{f}_{J_2}}=\cfrac{6J_2\mu_e R_e^2}{R_c^5}
    \begin{bmatrix}
        a^{(11)}_{J_2}& a^{(12)}_{J_2} & a^{(13)}_{J_2}\\
        a^{(21)}_{J_2}& a^{(22)}_{J_2} & a^{(23)}_{J_2}\\
        a^{(31)}_{J_2}& a^{(32)}_{J_2} & a^{(33)}_{J_2}\\
    \end{bmatrix}\boldsymbol{r},\label{fp_j2}
\end{equation}
where
\begin{equation}
\begin{aligned}
a^{(11)}_{J_2}&=1-3\sin^2{I}\sin^2{\theta},\quad
a^{(12)}_{J_2}=\sin^2{I} \sin{2\theta}, \\
a^{(13)}_{J_2}&=\sin{2I} \sin{\theta}, \quad
a^{(21)}_{J_2}=\sin^2{I}\sin{2\theta}, \\
a^{(22)}_{J_2}&=\sin^2{I}\left(\cfrac{7}{4}\sin^2{\theta}-\cfrac{1}{2}\right)-\cfrac{1}{4}, \quad
a^{(23)}_{J_2}=-\cfrac{1}{4}\sin{2I}\cos{\theta}, \\
a^{(31)}_{J_2}&=\sin{2I}\sin{\theta}, \quad
a^{(32)}_{J_2}=-\cfrac{1}{4}\sin{2I}\cos{\theta}, \\
a^{(33)}_{J_2}&=\sin^2{I}\left(\cfrac{5}{4}\sin^2{I}+\cfrac{1}{2}\right)-\cfrac{3}{4}. 
\end{aligned}
\end{equation}
A comparison of Eqs. (\ref{fc_j2}) and (\ref{fp_j2}) shows that $||{\boldsymbol{f}_{J_2}}||$ is always less than $||\boldsymbol{F}_{J_2}||$ under the assumption that $r \ll R_c$. 

The $J_2$-perturbed motion of the chief can also yield the relative fictitious accelerations of $\boldsymbol{f}_{o}$ and $\boldsymbol{f}_{f}$.  The osculating orbital elements owing to the $J_2$ potential for the near-circular orbit are obtained by substituting $e \xrightarrow{} 0$ into the known analytical solutions (e.g., see \citealt{Kozai1959motion, Brouwer1959solution}), as given in Appendix \ref{J2-perturbed orbital elements}. Notably, the $J_2$ perturbation produces a secular drift in the RAAN. This drift may be useful for the mission orbit to scan the visible sky area, as discussed in detail in Sect. \ref{Results and discussion}. 

\subsubsection{Earth $J_3$ gravity potential}
The Earth $J_3$ gravitational potential represents asymmetry with respect to the equatorial
plane. Its coefficient is on the order of $10^{-6}$, which is thee orders of magnitude smaller than that for the $J_2$ term; therefore, most effects on the $J_3$ term can be neglected in comparison to the $J_2$-perturbed effects. The only effect to be considered is the long-period $J_3$ perturbation of the eccentricity vector in the presence of the $J_2$ perturbation. 

Assuming a small eccentricity and neglecting longitude-dependent effects, Lagrange's equations relating to the eccentricity vector for the long-period $J_3$ perturbation are as follows \citep{Kozai1959motion}:
\begin{eqnarray}
\left\{
\begin{aligned}
    \dot{e}_{x}&=-\dot{\bar{\omega}}{e}_{y}+\cfrac{3J_3R_e^3{n} }{2a^3}\sin{I}\left(\cfrac{5}{4}\sin^2{I}-1\right),\\
    \dot{e}_{y}&=\dot{\bar{\omega}}{e}_{x},\label{lpe_j3}
\end{aligned}
\right.
\end{eqnarray}
where $\dot{\bar{\omega}}$ is the $J_2$-perturbed secular drift rate of the argument of perigee given by $\dot{\bar{\omega}}=(3/4)J_2(R_e/a)^2(5\cos^2{I}-1)n$ for the near-circular orbit. By applying the equilibrium condition of the eccentricity vector to an arbitrary inclination, the equilibrium eccentricity vector, known as a frozen eccentricity vector $(\boldsymbol{e}_f)$, is given as follows \citep{Kozai1959motion}: 
\begin{equation}
    \boldsymbol{e}_f=
    \begin{bmatrix}
        0\\
        -\cfrac{J_3 R_e}{2J_2a}\sin{I}
    \end{bmatrix}.\label{e_frozen}
\end{equation}
Equation (\ref{e_frozen}) implies that the mean eccentricity vector cannot be maintained at zero under the existence of Earth's $J_2$ and $J_3$ gravity potentials, and its effect can be significant, particularly in LEO. For example, for a satellite in a near-circular LEO at an altitude of 500 km and inclination of 100$^\circ$, the frozen eccentricity becomes $\boldsymbol{e}_f=[0,0.0011]^T$. Such a nonzero eccentricity vector can contribute to the relative fictitious accelerations of $\boldsymbol{f}_o$ and $\boldsymbol{f}_f$. 

\subsubsection{Lunisolar gravity}
The absolute lunisolar gravity acceleration vector of the chief with respect to the Earth ($\boldsymbol{F}_{3rd}$, as a part of $\boldsymbol{F}_c$) is approximated as \citep{Montenbruck2000}
\begin{equation} 
    \boldsymbol{F}_{3rd}={C}_{I}^{L}\left[\frac{\mu_b}{R_b^3}\left\{-\boldsymbol{R}_c^I+3\left(\cfrac{\boldsymbol{R}_c^I \cdot {\boldsymbol{R}_b^I}}{R_b^2}\right){\boldsymbol{R}_b^I} \right\} \right], \label{fc_3rd}
\end{equation}
where ${\boldsymbol{R}_b^I}$ is the position vector of the third body in the ECI frame, $\mu_b$ denotes the gravity constant of the third body, ${C}_{I}^{L}$ is the rotation matrix from the ECI frame to the rotational (LVLH) frame of the chief, and $R_b=||{\boldsymbol{R}_b^I}||$. 

The linearized gravity gradient acceleration vector owing to the third body (${\boldsymbol{f}_{3rd}}$, as a part of $\boldsymbol{f}_p$) is
\begin{equation}
    {\boldsymbol{f}_{3rd}}={C}_{I}^{L} \left[\frac{\mu_b}{R_b^3}\left\{-\boldsymbol{r}^I+3\left(\cfrac{\boldsymbol{r}^I \cdot {\boldsymbol{R}_b^I}}{R_b^2}\right){\boldsymbol{R}_b^I} \right\} \right]. \label{fp_3rd}
\end{equation}
Assume that $r \ll a$; then, $||{\boldsymbol{f}_{3rd}}||$ is always less than $||\boldsymbol{F}_{3rd}||$. 

The perturbed motion of the chief by lunisolar gravity can also yield the relative fictitious accelerations of $\boldsymbol{f}_{o}$ and $\boldsymbol{f}_{f}$; therefore, understanding its osculating motion is important. To the best of our knowledge, however, no appropriate analytical expressions describing the secular, long-periodic, and short-period motions in a near-circular geocentric orbit due to lunisolar gravity have been proposed with reasonable accuracy for a relatively short timescale. 
Early work \citep{Cook1962lunisolar} analytically calculated the secular and long-period satellite motion due to lunisolar perturbations up to the second degree of Legendre polynomials, but short-period motions were not analyzed. Subsequent work \citep{Kozai1973} analytically proposed the calculation of short-period satellite motion caused by lunisolar perturbations, but secular and long-period motions were calculated numerically. Recent work \citep{Roscoe2013third} developed secular and periodic motion models owing to third-body perturbations using a single- and double-averaged disturbing function. Analytical models were developed based on the assumption of a circular-restricted three-body problem; however, this may be problematic, particularly when long-term orbital motion on the eccentricity vector is a major concern. It is known that the parallactic term (third harmonic) of the lunar disturbing function can change the eccentricity of a circular geocentric orbit (e.g., see the Introduction in \citet{Allan1964long}) through the nonzero eccentricity of the lunar orbit. To capture this effect, the disturbing function of the third body is considered up to the third degree of the Legendre polynomials. Taking these points into account, this study analytically developed a relatively simple, near-circular satellite motion model based on to lunisolar gravity, and the results are given in Appendix \ref{Third-body-perturbed orbital elements}. 

In addition, the long-term effects of the lunisolar gravitational field and $J_2$ term of the Earth’s oblateness may be comparable for a medium-to-high Earth orbit. For this gravity field, equilibrium solutions exist in which the averaged chief orbits may be frozen. This plane is known as the classical Laplace plane, which is between the Earth's equator and that of the ecliptic, passing the vernal equinox \citep{Allan1964long, Tremaine_2009, ROSENGREN2014classical}. Assuming that the chief orbit is circular, the lunar orbit lies in the ecliptic, and the direction of the angular momentum for the chief lies on the principal plane defined by the Earth polar axis and normal direction of the Earth orbit (such that $\Omega=0^\circ$ or $180^\circ$), the frozen orbit condition is \citep{Allan1964long}
\begin{equation}
    \tan{\Phi}=\frac{(\omega_m+\omega_s)\sin{2\epsilon_e}}{\omega_{J_2}+(\omega_m+\omega_s)\cos{2 \epsilon_e}},\label{laplace_frozen}
\end{equation}
where $\epsilon_e$ is the obliquity of the ecliptic ($\epsilon_e=23.4^{\circ}$), $\Phi$ is the azimuth angle, and $\omega_{J_2}$, $\omega_m$, and $\omega_s$ are defined as
\begin{eqnarray}
\begin{aligned}
    \omega_{J_2}&=\frac{3nJ_2R_e^2}{2a^2},\\
    \omega_m&=\frac{3\mu_m}{4na_m(1-e_m^2)^{3/2}},\quad \omega_s=\frac{3\mu_s}{4na_s(1-e_s^2)^{3/2}}.
\end{aligned}
\end{eqnarray}
The subscripts $m$ and $s$ indicate the Moon or the Sun, respectively, and $a$ and $e$ with these subscripts represent the semimajor axis and eccentricity for the Moon or the Sun. For example, for a chief satellite at an altitude of 35,800 km (geosynchronous) with $\Omega=0^{\circ}$, one can find that $I=\Phi=7.4^{\circ}$ is one of the solutions for the frozen orbits in Eq. (\ref{laplace_frozen}). If the circular orbit does not lie in the classical Laplace plane, the orbital orientation will oscillate around the frozen point with a period of up to several tens of years. The frozen orbit lying in the classical Laplace plane may be useful particularly for the gravitational-wave detector, for obtaining the long-term stability on the orbital orientation with respect to the Sun, as discussed in Sect. \ref{Gravitational-wave telescope}.

\subsubsection{Atmospheric drag}
Atmospheric drag is a small force for most orbits, but it can be a major perturbation source for relative motion in LEOs. The atmospheric-drag acceleration vector acting on the chief spacecraft in the LVLH frame ($\boldsymbol{F}_D$, as a part of $\boldsymbol{F}_c$) is as follows:
\begin{equation}
    \boldsymbol{F}_{D}=-\frac{1}{2}\rho V^2 B_D \left(\cfrac{\boldsymbol{V}}{V}\right)=-\frac{\mu_e}{2a}\rho B_D \left(\cfrac{\boldsymbol{V}}{V}\right),\label{fc_drag}
\end{equation}
where $\boldsymbol{V}$ is the velocity vector of the chief for the unperturbed and circular assumption in the LVLH frame ($V=||\boldsymbol{V}||{=na}$), $\rho$ is atmospheric density, $B_D(=S C_D/m)$ is the coefficient for drag, $S$ is the cross-sectional area, $C_D$ is the drag coefficient, and $m$ is the spacecraft mass. Equation (\ref{fc_drag}) is obtained by assuming that the atmosphere co-rotates with the Earth \citep{Montenbruck2000} and $\boldsymbol{V}$ is approximately one order of magnitude higher than that of the Earth-fixed atmosphere; thus, atmospheric velocity is neglected. 

In general, the atmospheric density has a large uncertainty, mainly owing to diurnal variation, periodic solar activity, and sudden solar activity \citep{Montenbruck2000}. However, the most evident dependence of atmospheric density is that it decreases with increasing altitude. The Harris--Priester density model \citep{Harris1962} is relatively simple and is widely used as a standard atmospheric model. The model is given as follows: 
\begin{equation}
    \rho(h)=\rho_m(h)+\left[\rho_M(h)-\rho_m(h)\right]\cos^N{\left(\cfrac{\Psi}{2}\right)},
\end{equation}
where $h$ is the height above Earth's reference ellipsoid, $\Psi$ is the angle between the satellite position vector and the apex of the diurnal bulge, $N$ is a numerical value set to 2 for low-inclination orbits and 6 for polar orbits, and $\rho_M(h)$ and $\rho_m(h)$ are the apex and antapex densities, respectively. It should be noted that $\rho(h)$ can be directly computed from the table of Harris--Priester atmospheric-density coefficients for mean solar activity \citep{Long1989}. When assuming a spherical Earth, height is given by $h=(a-R_e)$; thus, $\rho$ is written as a function of the semimajor axis. 

Assuming that the two satellites are flying in proximity formation, such that $V$ and $\rho$ are the same for both spacecraft, the relative acceleration vector for the deputy caused by atmospheric drag ({$\boldsymbol{f}_{D}$, as a part of $\boldsymbol{f}_p$}) can be modeled as follows:
\begin{equation}
{\boldsymbol{f}_{D}=-\frac{\mu_e}{2a}\rho (\delta B_D)} \left(\cfrac{\boldsymbol{V}}{V}\right),\label{fp_drag}
\end{equation}
where $\delta B_D$ is the difference between the coefficients $B_D$ of the two spacecraft for atmospheric drag. 

In contrast to Earth $J_2$ and lunisolar gravity, $||\boldsymbol{F}_{D}||$ and $||{\boldsymbol{f}_{D}}||$ can be of the same order. In addition, drag is a small force in most Earth orbits. Therefore, we assumed that both $\boldsymbol{F}_{D}$ and ${\boldsymbol{f}_{D}}$ are compensated for by the control accelerations, and we neglected the effects of $\boldsymbol{f}_{o}$ and $\boldsymbol{f}_{f}$.  

\subsubsection{Solar radiation pressure}
Solar radiation pressure (SRP) can be regarded as a function of the Sun-to-spacecraft distance, such that its magnitude is almost identical for spacecraft orbiting the Earth. A cannonball model was used, in which the satellite shape was assumed to be a uniform sphere. When the shade effect of the Earth and the small eccentricity of the Earth's orbit are neglected, the acceleration vector by SRP acting on the chief spacecraft in the LVLH frame ($\boldsymbol{F}_s$, as a part of $\boldsymbol{F}_c$) is expressed as follows \citep{Montenbruck2000}:
\begin{equation}
    \boldsymbol{F}_s =-P_s \left(\frac{C_s S}{m}\right)\boldsymbol{s} =-P_s B_s \boldsymbol{s},\label{fc_srp}
\end{equation}
where $P_s$ is the SRP, $C_s$ is the radiation pressure coefficient, $B_s$ is the coefficient for SRP, and $\boldsymbol{s}$ is the unit vector pointing to the Sun in the LVLH frame. Assuming that two spacecraft are flying in proximity formation ($P_s$ is the same for both spacecraft), the relative acceleration vector for the deputy caused by SRP ($\boldsymbol{f}_s$, as a part of $\boldsymbol{f}_p$) can be modeled as follows \citep{koenig2017new}:
\begin{equation}
{\boldsymbol{f}_s} =-P_s {(\delta B_s)} \boldsymbol{s},\label{fp_srp}
\end{equation}
where $\delta B_s$ is the difference between the coefficients $B_s$ of the two spacecraft for SRP. Similar to atmospheric drag, $||\boldsymbol{F}_{s}||$ and $||{\boldsymbol{f}_s}||$ can be of the same order and their acceleration magnitudes are small. Therefore, we assumed that both $\boldsymbol{F}_{s}$ and ${\boldsymbol{f}_s}$ are compensated for by the control accelerations and neglected and the effects of $\boldsymbol{f}_{o}$ and $\boldsymbol{f}_{f}$.

\subsection{Control approach}\label{Control approach}

The orbit and formation control approach used in this study is determined based on the characteristics of each perturbation. Table \ref{table3} summarizes the control approach used to mitigate each perturbation. The check mark in Table \ref{table3} indicates that all marked perturbations are compensated for by the control, and the check mark with an asterisk indicates that either the absolute or relative perturbations are compensated for. As summarized in Table \ref{table3}, the CW nonlinearity must be nullified by the control to maintain the rigid formation; compensating for the relative perturbations for the Earth $J_2$ gravity potential and lunisolar gravity is economical (except for the secular and long-period effects on the eccentricity vector); the control strategy against the secular and long-period changes on the eccentricity vector owing to the lunisolar and Earth $J_3$ gravity potential are compared to decide whether to compensate for the absolute or relative perturbations; the absolute and relative accelerations for the atmospheric drag and SRP, both of which are small in most Earth orbits, are corrected by the control (although the compensation of the absolute accelerations may be unnecessary for certain applications). 

\begin{table}
\caption{\label{table3} Control approach for mitigating each perturbation in this study. }
\centering
\begin{tabular}{lllll}
\hline \hline 
Perturbations & $\boldsymbol{F}_c$ & $\boldsymbol{f}_p$ & $\boldsymbol{f}_o$ & $\boldsymbol{f}_f$\\\hline
CW nonlinearity & &\checkmark & & \\
Earth $J_2$ gravity &  &\checkmark & \checkmark & \checkmark \\
Earth $J_3$ gravity ($e^{(lp)}$) & $\checkmark^*$ & & $\checkmark^*$ & $\checkmark^*$ \\
Lunisolar gravity &  &\checkmark & \checkmark & \checkmark \\
Lunisolar gravity ($e^{(s)}$) & $\checkmark^*$ & & $\checkmark^*$ & $\checkmark^*$ \\
Lunisolar gravity ($e^{(lp)}$) & $\checkmark^*$ & & $\checkmark^*$ & $\checkmark^*$ \\
Atmospheric drag & \checkmark &\checkmark &  & \\
Solar radiation pressure & \checkmark &\checkmark &  &\\  \hline
\end{tabular}
\tablefoot{
A check mark indicates that all marked perturbations are compensated for by the control, and a check mark with an asterisk indicates that either the absolute or relative perturbations are compensated for. The perturbation specified as ``Lunisolar gravity'' excludes its secular and long-period effects on the eccentricity vector drift. The $e^{(s)}$ and $e^{(lp)}$ indicate the secular and long-period perturbations on the eccentricity vector, respectively. }
\end{table}

\begin{figure*}
        \centering
        \resizebox{\hsize}{!}{\includegraphics[width=1\columnwidth]{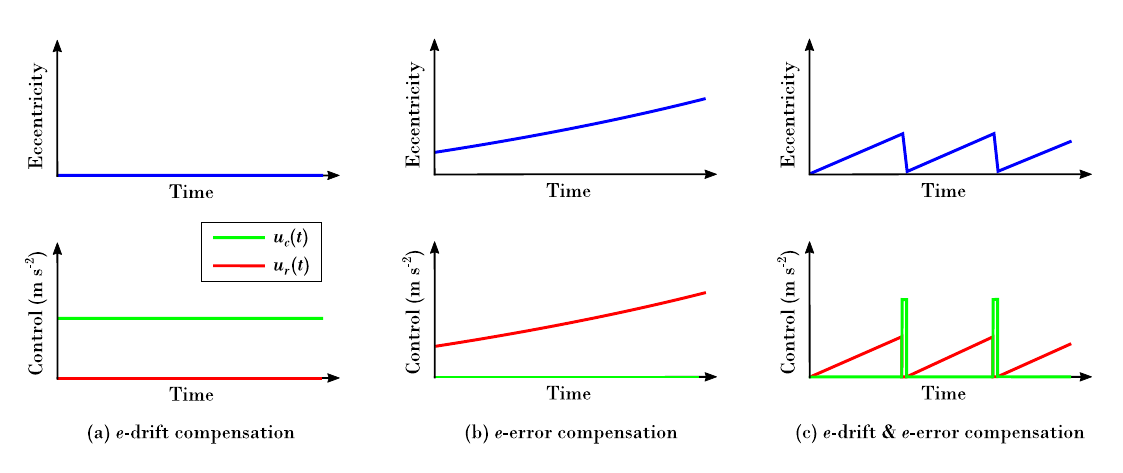}}        
        \caption{Three methods for mitigating the secular and long-period eccentricity vector drift. The panels on the left ((a) $e$-drift compensation) show the approach for compensating for the eccentricity vector drift of the chief using continuous control. The center panels ((b) $e$-error compensation) show the method that accepts the small eccentricity error and compensates for the relative perturbing acceleration of the deputy via continuous control. The panels on the right ((c) $e$-drift and $e$-error compensation) show the hybrid approach. The top and bottom panels for each approach show the time histories of the eccentricity and control accelerations, respectively. }
        \label{fig1}
\end{figure*}

We propose three possible methods for mitigating the secular and long-period effects on the eccentricity vector, as shown in Fig. \ref{fig1}. The first approach is to maintain the eccentricity vector at zero by nullifying secular and/or long-period drifts using continuous control. For a near-circular orbit, the radial control force ($u_x$) is sensitive to the eccentricity vector and mean argument of latitude; the tangential force ($u_y$) is sensitive to the eccentricity vector and semimajor axis; and the normal force ($u_z$) is sensitive to the inclination and RAAN. Therefore, when $u_x$ and $u_y$ are used to compensate for the eccentricity vector drift, Gauss's variational equations for the eccentricity vector are as follows:
\begin{eqnarray}
\left\{
\begin{aligned}
    \dot{\hat{e}}_{x}&= \cfrac{1}{n a}(u_x\sin{\phi}+2u_y\cos{\phi}),\\
    \dot{\hat{e}}_{y}&= \cfrac{1}{n a}(-u_x\cos{\phi}+2u_y\sin{\phi}),
\end{aligned}
\right.
\end{eqnarray}
where $\dot{\hat{e}}_{x}$ and $\dot{\hat{e}}_{y}$ denote the change rates of the eccentricity vector by the control. Terms with $e$ and $e^2$ are neglected in these equations. Setting $u_x$ and $u_y$ as
\begin{eqnarray}
\left\{
\begin{aligned}
    u_x&= n a|\dot{e}|\sin{(\phi+\phi_0)},\\
    u_y&=\cfrac{1}{2}n a|\dot{e}|\cos{(\phi+\phi_0)},
\end{aligned}
\right.
\label{lpe_cont}
\end{eqnarray}
provides the necessary control to compensate for the eccentricity vector drift, where $|\dot{e}|=\sqrt{\dot{e}^2_{x}+\dot{e}^2_{y}}$, $\cos{\phi_0}=-\dot{e}_{x}/|\dot{e}|$, and $\sin{\phi_0}=\dot{e}_{y}/|\dot{e}|$. In this context, $\dot{e}_{x}$ and $\dot{e}_{y}$ are considered as the time derivatives of the secular and long-period variations on the eccentricity vector without control. This control law does not yield the secular and long-period changes in $a$ and $\phi$. The second approach is to accept the small eccentricity and compensate for its relative perturbing acceleration using continuous control. As previously mentioned, the control acceleration can be calculated using Eqs. (\ref{F_o}) and (\ref{F_f2}). The third approach combines the previous two methods: the eccentricity vector drift is left as long as it remains within an acceptable level, and the accumulated eccentricity is periodically nullified by impulsive or continuous control. A benefit of the third approach is that the propellant for the observation may be saved by using another propulsion system for the satellite bus system (if available) because the objective of this control is only to reconfigure the orbit of the chief, whereas the previous two methods must realize the formation maintenance for the observation and orbit control of the chief simultaneously; thus, the propellant for the observation cannot be saved. 

For long-period perturbations owing to $J_3$ and lunisolar gravity, the eccentricity vector can remain around the frozen point. Thus, the first and second methods are compared. For secular perturbations due to lunisolar gravity, the eccentricity can increase linearly with time; thus, the secular drift of the eccentricity vector may not be acceptable. Thus, the first and third methods are compared. 

\subsection{Orbit selection approach}\label{Orbit selection approach}

By analyzing the factors that contribute the most to the control magnitudes, further insight is attained regarding the orbit selection approach. The developed perturbation models suggest that the semimajor axis ($a$) characterizes the absolute perturbations, whereas the semimajor axis and formation size ($r_r$) characterize the relative perturbations. Therefore, the orders of the control magnitudes for $\boldsymbol{F}_c$ and {$\boldsymbol{f}_p$} can be evaluated by an explicit analysis of their accelerations as a function of $a$ and/or $r_r$. Similarly, the control magnitudes for {$\boldsymbol{f}_o$} and {$\boldsymbol{f}_f$} can be evaluated using a two-step analysis to first evaluate the factors for the perturbed orbital elements of the chief and then calculate the orders of their perturbing accelerations as a function of $a$ and $r_r$. The three types of relative perturbations ({$\boldsymbol{f}_p$, $\boldsymbol{f}_o$, $\boldsymbol{f}_f$}, if available) have the same order against each perturbation source provided that $\rho_x$, $\rho_y$, and $\rho_z$ are derived from the natural solution in Eq. (\ref{CW_bounded_solution}). 

Table \ref{table4} summarizes the factors that contribute the most to the control magnitudes for mitigating the absolute and relative perturbations, characterized by $a$, $r_r$, and other important parameters. The control magnitudes against the absolute perturbations relating to Earth’s $J_2$ gravity, the long-period drift of the eccentricity vector owing to Earth’s $J_3$ gravity, and atmospheric drag decrease with an increase in $a$, whereas those of the lunisolar gravity increase, and that of the SRP remains constant. The control magnitudes against the relative perturbations for Earth’s $J_2$ gravity, long-period drift of the eccentricity vector owing to Earth’s $J_3$ gravity, atmospheric drag, and SRP maintain the same trend as that for the absolute perturbations. Interestingly, those of the lunisolar gravity (except for the long-term drift of the eccentricity vector) remain constant, and their secular and long-period effects on the eccentricity even decrease with an increase in $a$. Regarding the relationship with $r_r$, the control magnitudes against the relative perturbations increase linearly with $r_r$ for Earth’s $J_2$ gravity, the long-period drift of the eccentricity vector owing to Earth’s $J_3$ gravity, and lunisolar gravity, and are squared with $r_r$ for the CW nonlinearity. In contrast, those for the atmospheric drag and SRP remain constant with respect to $r_r$. 

To mitigate the long-term drift on the eccentricity vector, a more propellant-saving approach for Earth’s $J_3$ gravity is the compensation for the absolute perturbations (with a factor of $a^{-5}$) in the larger semimajor axis and size of formation, and the relative perturbations (with a factor of $a^{-4} r_r$) for smaller $a$ and $r_r$. However, a better approach for the lunisolar gravity is to compensate for the absolute perturbations in the smaller $a$ and larger $r_r$ as well as the relative perturbations in the larger $a$ and smaller $r_r$. 

Overall, small-perturbation regions tend to appear in the higher-altitude and shorter-separation regions. Thus, a reasonable approach for orbit selection is to search for the candidates of the semimajor axis that satisfy the small-disturbance conditions under the desired formation size. However, the orbital orientation parameters ($I$ and $\Omega$) are selected to satisfy the observation conditions, considering the long-term perturbations. Because the observation conditions are highly mission-oriented, the orbit selection approach is verified through case studies in the next section. 

\begin{table}
\caption{Largest contributing factors to the control magnitudes. }\label{table4} 
\centering
\begin{tabular}{lcc}
\hline \hline 
Perturbations & \multicolumn{2}{c}{Contributing factors}\\
 & Absolute & Relative \\\hline
CW nonlinearity & N/A & $a^{-4}r_r^2$\\
Earth $J_2$ gravity &$a^{-4}$ & $a^{-5}r_r$\\
Earth $J_3$ gravity ($e^{(lp)}$)  & $a^{-5}$ & $a^{-4}r_r$ \\
Lunisolar gravity & $a$ & $r_r$\\
Lunisolar gravity ($e^{(s)}$) & $a^{2}$ & $a^{-1/2}r_rt$ \\
Lunisolar gravity ($e^{(lp)}$) & $a^{2}$ & $a^{-1/2}r_r$ \\ 
Atmospheric drag & $\rho a^{-1}$ & $\rho a^{-1}$ \\
Solar radiation pressure & Const. & Const. \\  \hline
\end{tabular}
\tablefoot{For Earth $J_3$ gravity ($e^{(lp)}$) and lunisolar gravity ($e^{(s)}$ and $e^{(lp)}$), the eccentricity originally had a contributing factor of $(en^2r_r) \propto (ea^{-3}r_r)$ to the relative fictitious accelerations, as shown in Sect. \ref{Perturbation periods and acceleration magnitudes}. Subsequently, the contributing factors were calculated by substituting the secular and long-period parts of the eccentricity for Earth $J_3$ and lunisolar gravity into $(ea^{-3}r_r)$. }
\end{table}

\section{Results and discussion}\label{Results and discussion}
Numerical examples of control acceleration magnitudes, observabilities, and visibilities are provided to verify the proposed approach to formation-flying interferometry. The selected case studies were a laser-interferometric gravitational-wave telescope and an astronomical interferometer. 

The analytical perturbation models developed in this study were used to analyze the control acceleration magnitudes and observation conditions. In addition, the control acceleration magnitudes were evaluated by averaging each perturbing acceleration per orbital revolution from the initial time and then calculating their root-sum-square (Rss). Table \ref{table5} lists the common numerical conditions for all formation configurations. The area-to-mass ratio was assumed for a small class of satellites (e.g., $M\approx500$ kg and $S\approx2.3$ $\rm m^2$), and its value was used to calculate atmospheric drag and SRP effects. To calculate the control acceleration by method (c) in Fig. \ref{fig1}, the secular drift of the eccentricity vector owing to lunisolar gravity was assumed to be periodically corrected to zero each year. The propellant consumption for this correction did not consider the control acceleration; the consumption to mitigate the relative acceleration due to the mean (nonzero) eccentricity vector accumulated over one year was considered only. The study did not account for propellant consumption related to formation reconfiguration maneuvers, such as altering observation directions or the size of the formation. 

In the heliocentric orbit or halo orbit around the Sun-Earth L2 point, SRP is the dominant disturbance. Using the values in Table \ref{table5}, the sum of the absolute and relative perturbing accelerations owing to SRP was calculated as approximately $4.5\times10^{-8}$ $\rm m\,s^{-2}$ (1.4 $\rm m\,s^{-1}$ per year). Therefore, this study aims to find a small disturbance environment in geocentric orbits of less than $1.0\times10^{-7}$ $\rm m\,s^{-2}$ (3.0 $\rm m\,s^{-1}$ per year) that maintains a magnitude only a few times larger than that in the heliocentric or Sun-Earth L2 orbits. 

\begin{table}
\caption{Typical numerical conditions for control-acceleration analysis.}\label{table5} 
\centering
\begin{tabular}{lll}
\hline \hline 
Item & Symbol & Conditions \\ \hline
Initial time & - & 2024 January 1, \\ 
 &  & 00:00:00 UT \\ 
Initial orbital elements & - & $\boldsymbol{e}_{0}^{\rm (s)}=\boldsymbol{0}$, $\phi_0^{\rm(s)}=0$ \\ 
Area-to-mass ratio& $(S/m)$ & $4.5 \times 10^{-3}$ $\rm m^2\,kg^{-1}$  \\ 
Atmospheric drag  & $C_d$& 2.5 \\ 
  & $(\delta B_d/B_d)$ & 0.1 \\ 
Solar radiation pressure & $P_s$ & $4.56\times10^{-6}$ Pa \\ 
& $C_s$ & 2 \\ 
& $(\delta B_s/B_s)$ & 0.1 \\ 
Control period ($e^{\rm(s)}$) & - & 1 year \\ \hline
\end{tabular}
\tablefoot{The initial orbital elements of $a_{0}^{\rm (s)}$, $I_{0}^{\rm (s)}$, and $\Omega_{0}^{\rm (s)}$ were selected based on each mission application, as given in Sects. \ref{Gravitational-wave telescope}--\ref{LEO environment}. }
\end{table}

\subsection{Gravitational-wave telescope}\label{Gravitational-wave telescope}
The assumed laser-interferometric gravitational-wave telescope in this study is B-DECIGO \citep{kawamura2021current}, which is a precursor to DECIGO. B-DECIGO has three pairs of Fabry-P\'{e}rot cavities, and one cluster of three satellites is placed equidistantly, 100 km apart, in a triangular formation. 

\begin{figure}
        \centering
        \includegraphics[width=1\columnwidth]{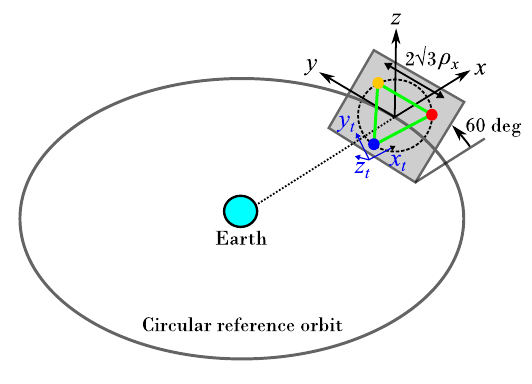} 
        \caption{Reference formation orbit for the laser-interferometric gravitational-wave observatory. The distance between each spacecraft is given by $2\sqrt{3}\rho_x$ (constant). }\label{fig2}
\end{figure}

The triangular reference formation for B-DECIGO uses a general circular orbit \citep[GCO;][]{Alfriend2010spacecraft}, which is achieved by setting the parameters in Eq. (\ref{CW_bounded_solution}) as $\rho_y=0$, $\rho_z=\sqrt{3}\rho_x$, and $\alpha_z=\alpha_x$.  The GCO is always inclined by $60^{\circ}$ with respect to the orbital plane of the chief, because $\tan^{-1}{(z_r/\sqrt{x_r^2+y_r^2})}=\pi/3$, where $\boldsymbol{r}_r=[x_r,\,y_r,\,z_r]^T$. 
The reference formation can be configured by placing the three spacecraft in a GCO with orbit phase angles of $2\pi/3$ from each other. The radius of the GCO ($r_r$) and the distance between each spacecraft ($L_{12}$, $L_{23}$, $L_{31}$) remain constant such that $r_{r}=2\rho_x$ and $L_{12}=L_{23}=L_{31}=2\sqrt{3}\rho_x$. The control requirement for B-DECIGO is imposed on the stability of each optical path length between the mirrors comprising the Fabry-P\'{e}rot cavity (e.g., $\int \dot{L}_{12}(t)dt$) within the order of nanometers. Based on the B-DECIGO requirements, the sizing parameter was selected as $r_r=100/\sqrt{3}$ km. Spacecraft 1 (SC1), 2 (SC2), and 3 (SC3) were assigned $\alpha_x=0$, $2\pi/3$, and $4\pi/3$, respectively. Figure \ref{fig2} illustrates the triangular formation of the GCO. For this reference formation, the first and second time derivatives of $\rho_x$, $\rho_y$, and $\rho_z$ are zero, but $\dot{\omega}_t$ is nonzero. Therefore, only the second term in Eq. (\ref{F_f2}) is nonzero, and the other terms are zero. 

The most important observation requirement for the orbital orientation of B-DECIGO is sunlight avoidance of the optical paths. The minimum avoidance angle was set to 15$^{\circ}$ in this study. This requirement can always be satisfied in a heliocentric orbit, whereas it can be violated in Earth orbits; therefore, a careful choice of orbital orientation parameters is necessary. To satisfy the sunlight-avoidance requirement, the parameters were selected as $I^{(s)}_0=20^{\circ}$ and $\Omega^{(s)}_0=0^{\circ}$ in a high Earth orbit for the following reasons. 

\begin{figure}
        \centering
        \includegraphics[width=1\columnwidth]{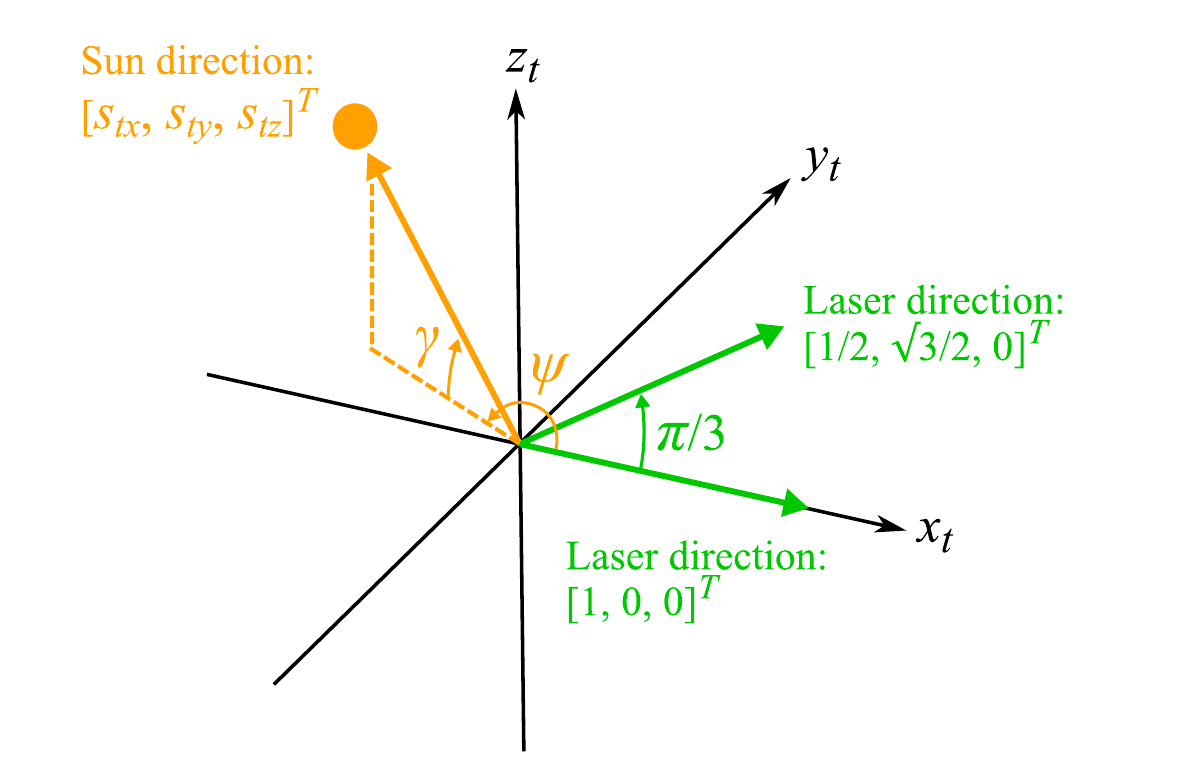} 
        \caption{Sun and laser directions in the target attitude frame. The azimuth and elevation angles ($\psi$, $\gamma$) of the Sun vector are also depicted. }\label{fig3}
\end{figure}

\begin{figure}
        \centering
        \includegraphics[width=1\columnwidth]{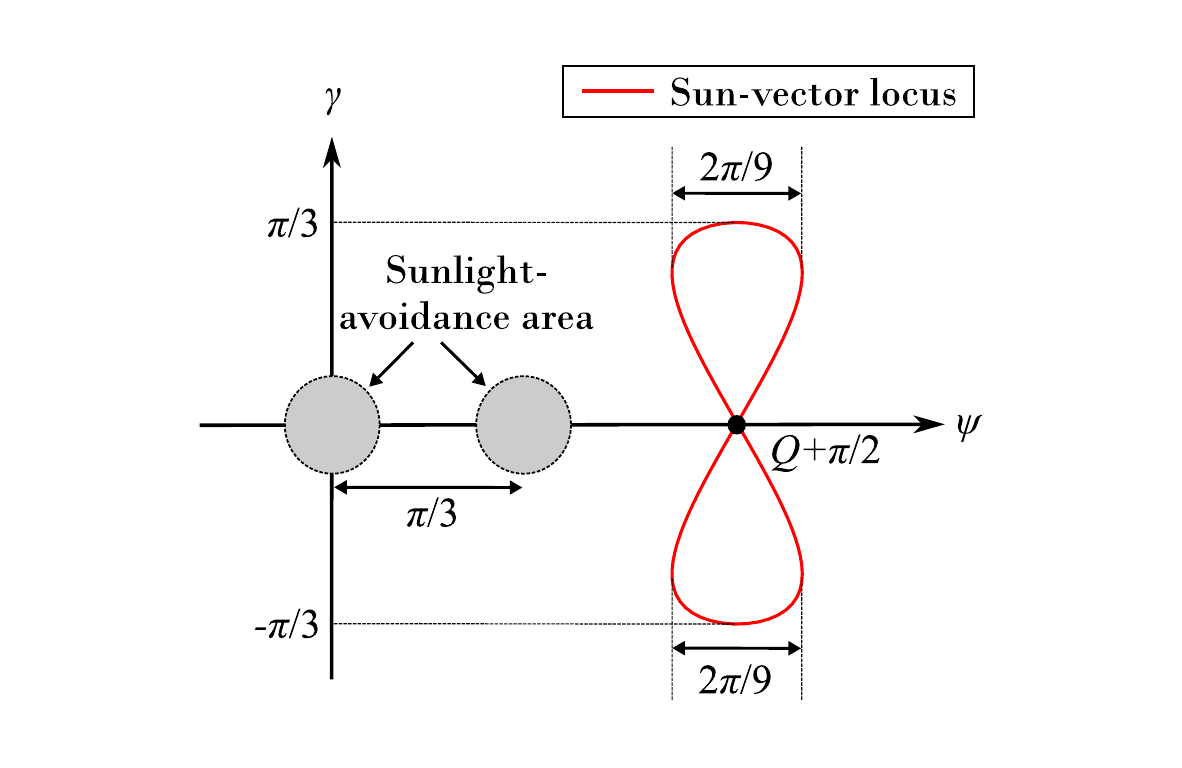} 
        \caption{Sunlight-avoidance area and Sun-vector locus per orbit with respect to the azimuth and elevation. The sunlight-avoidance areas for the two laser axes of each spacecraft are depicted as gray circles. Equation (\ref{sun_locus}) indicates that the Sun-vector locus per orbit has azimuth and elevation amplitudes of nearly $\pi/9$ and $\pi/3$, respectively. The angle $Q$ drifts by $2\pi$ over one year. }\label{fig4}
\end{figure}

The target attitude frame is illustrated in Figs. \ref{fig2} and \ref{fig3} such that the two laser beams of each spacecraft point to $[x_t, \, y_t, \, z_t]^T=[1, \, 0, \, 0]^T$ and $[1/2, \, \sqrt{3}/2, \, 0]^T$ and the $z_t$ axis is selected in the right-hand frame. In addition, by approximating the inclination as the same as the obliquity of the ecliptic, the direction of the Sun is constrained within the $x$-$y$ plane in the LVLH frame. Then, the Sun vector in the target attitude frame ($\boldsymbol{s}^t=[s_{tx},s_{ty},s_{tz}]^T$) is expressed as follows: 
\begin{equation}
\boldsymbol{s}^t=\cfrac{1}{4}
\begin{bmatrix}
     \sin{(2P+Q)}-3\sin{Q}\\
    -\cos{(2P+Q)}+3\cos{Q}\\
    -2\sqrt{3}\cos{P}
\end{bmatrix},\label{sun_locus}
\end{equation}
where
\begin{eqnarray}
P=\theta-\theta_s, \quad Q=\theta_s+\alpha_x+\cfrac{\pi}{6}, 
\end{eqnarray}
and $\theta_s$ is the argument of latitude of the Sun, changing by $2\pi$ over one year. By defining the azimuth angle ($\psi$) and elevation angle ($\gamma$) with respect to the Sun vector as $\psi={\rm arctan2}{(s_{ty}, s_{tx})}$ and $\gamma=\sin^{-1}{(s_{tz})}$, respectively (see Fig. \ref{fig3}), a pair of ($\psi$, $\gamma$) follows the locus similar to a ``figure of eight'' per orbit revolution of satellites passing the cross-point $(\psi,\, \gamma)=(Q+\pi/2,\,0)$ twice, as shown in Fig. \ref{fig4}. In addition, angle $Q$ drifts by $2\pi$ annually. Because the angle between the Sun vector and laser direction approaches zero when $(\psi,\,\gamma)\xrightarrow{}(0,\,0)$ or $(\pi/3,\,0)$ (see Fig. \ref{fig3}), each of the two laser axes in a single satellite is subject to interference by sunlight only once per year. Hence, excellent observability is expected. Furthermore, as the selected orbital plane approximately corresponds to the classical Laplace plane, long-term stability of the orbital orientation is expected, and this stability will maintain excellent observability.  

\begin{figure*}
        \centering
        \resizebox{\hsize}{!}{\includegraphics[width=1\columnwidth]{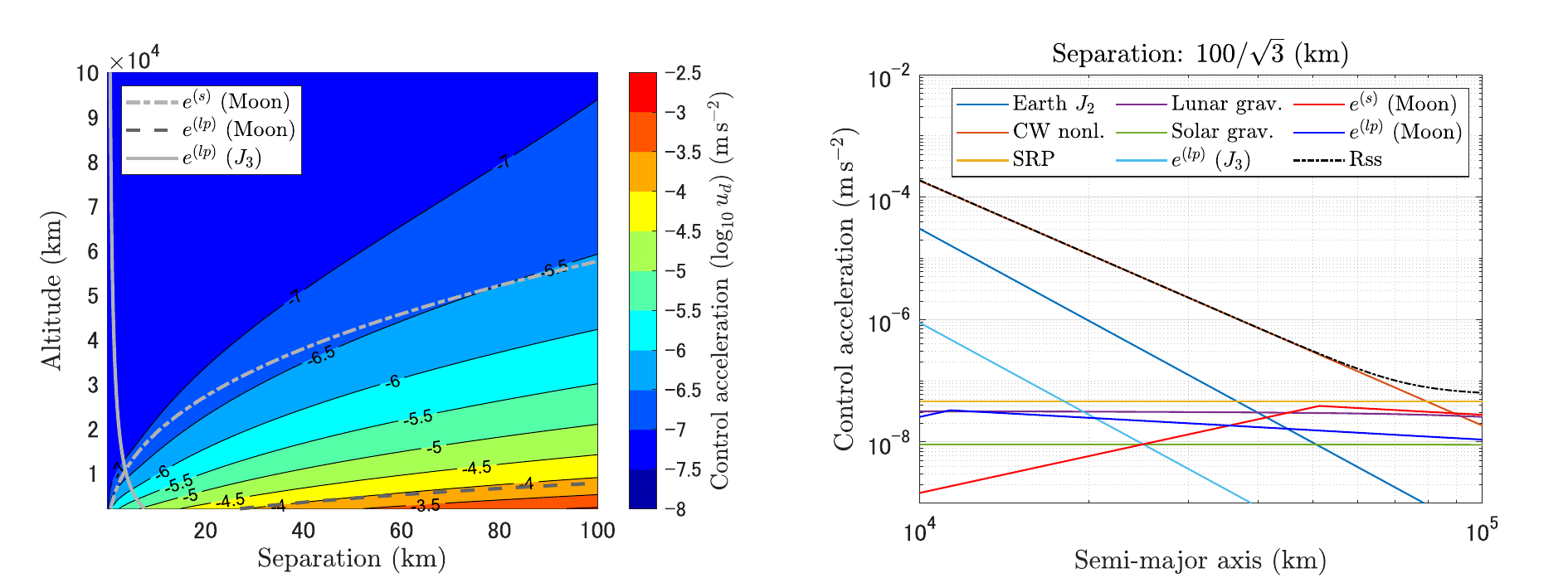}}        
    \caption{Control acceleration magnitudes for the laser-interferometric gravitational-wave observatory. The left panel shows the contour map of the control acceleration magnitudes ($u_d=||\boldsymbol{u}_d||$) with respect to the separation (defined as $r_r$) and altitude. The gray lines show the borderlines that indicate when to change the mitigation methods for the secular and long-period drifts of the eccentricity vector. 
    For the secular and long-period eccentricity drift owing to lunar gravity, the more propellant-saving approach appeared above the lines (specified by $e^{(s)}$(Moon) and $e^{(lp)}$(Moon)) when applying method (c) for the secular one and (b) for the long-period one in Fig. \ref{fig1}, while it appeared below the lines when applying method (a) in Fig. \ref{fig1}. For the long-period eccentricity drift owing to Earth’s $J_3$ gravity, the more propellant-saving approach appeared above the line (specified by $e^{(lp)}(J_3)$) when applying method (a) in Fig. \ref{fig1}, while it appeared below the line when applying method (b) in Fig. \ref{fig1}. 
    The borderlines for the secular and long-period drifts of the eccentricity vector owing to solar gravity were abbreviated because their magnitudes were sufficiently smaller than $10^{-9}$ $\rm m\,s^{-2}$ in the plotted region. The right panel shows the control acceleration magnitudes at the desired separation ($100/\sqrt{3}$ km) with respect to the semimajor axis. The atmospheric drag and secular and long-period drift of the eccentricity vector owing to solar gravity were removed from the figure because their magnitudes were sufficiently smaller than $10^{-9}$ $\rm m\,s^{-2}$. } \label{fig5}
\end{figure*}

\begin{figure}
        \centering
        \includegraphics[width=1\columnwidth]{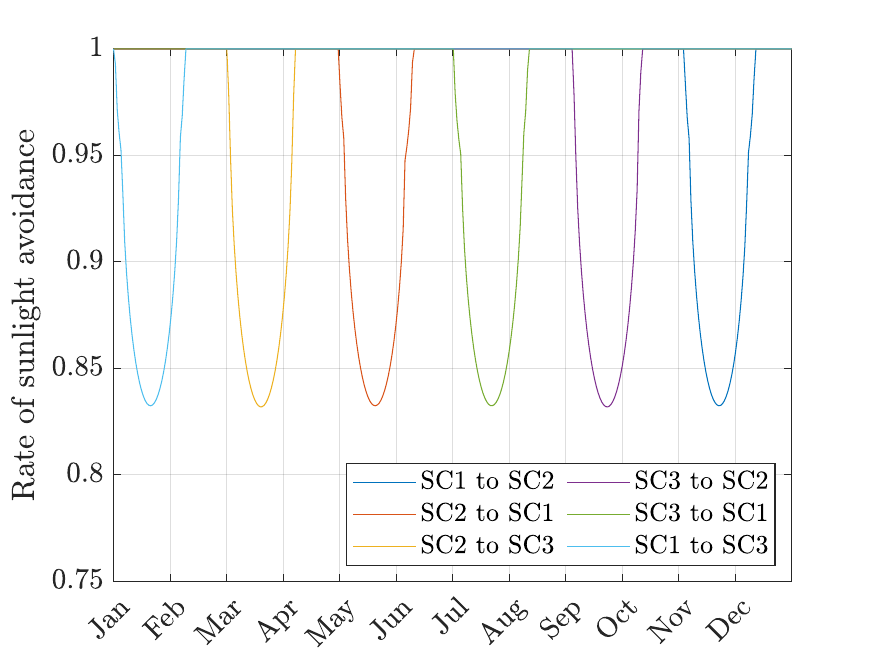}
        \caption{Rate of sunlight avoidance per orbit during one year (January 1 -- December 31, 2024). The six lines indicate the rates for the six laser-target directions: from SC1 to SC2, from SC2 to SC3, from SC3 to SC1, and their opposite directions. Each laser direction is subject to interference by the Sun once per year; thus, each cavity is interfered twice per year. } \label{fig6}
\end{figure}

To comprehensively understand the nature of geocentric orbit regions, the control acceleration magnitudes were computed over a wider range of spacecraft separations than the assumed value for B-DECIGO. The left panel of Fig. \ref{fig5} shows a contour map of the control acceleration magnitudes with respect to separation and altitude. The right panel shows the control acceleration magnitudes for each perturbing source at the desired separation ($100/\sqrt{3}$ km) for B-DECIGO with respect to the semimajor axis. 

The first feature is shown in Fig. \ref{fig5} left, indicating that the smaller control acceleration magnitudes appeared in the shorter-separation and higher-altitude region. Therefore, a small-disturbance region of less than $10^{-7}$ $\rm m\,s^{-2}$ was identified in high Earth orbit. The second feature is observed at the borderlines in Fig. \ref{fig5} left to attain more propellant-efficient control. In particular, the control approach to the secular drift of the eccentricity vector owing to lunar gravity switched from the mitigation of the absolute perturbing acceleration (in Fig. \ref{fig1} (a) with a factor of $a^2$) at an altitude lower than 51,000 km to that of the relative perturbing acceleration (in Fig. \ref{fig1} (c) with a factor of $a^{-1/2}rt$). Consequently, the total control acceleration magnitudes were maintained as low for $10^{5}$ km along the semimajor axis, as shown in Fig. \ref{fig5} right. 

By selecting the candidate Earth orbit for B-DECIGO at an altitude of 80,000 km ($a_0^{\rm (s)}=86,378$ km), the magnitude of the control acceleration was $7\times 10^{-8}$ $\rm m\,s^{-2}$. This acceleration can be further broken down to compensate for the absolute and relative perturbations, which account for 17 \% of the total acceleration ($||\boldsymbol{F}_c||$), 20 \% ($||\boldsymbol{f}_p||$), 33 \% ($||\boldsymbol{f}_o||$), and 30 \% ($||\boldsymbol{f}_f||$). This result suggests that these four types of perturbed acceleration may contribute in the same order, and all of them must be considered in the control approach. The eccentricity (initially zero) in the selected orbital environment can increase to $0.0012$ in a one-year duration owing to the Moon’s gravity, based on the satellite motion model in Appendix \ref{Third-body-perturbed orbital elements}. Thus, the eccentricity error must be corrected periodically.

Then, the observation conditions were analyzed at an altitude of 80,000 km. Figure \ref{fig6} shows the rate of sunlight avoidance per orbit for one year. Each of the three laser cavities interfered twice per year, and the interference season lasted for approximately 40 days. The minimum satisfaction rate was 83\% per orbit, which is approximately equivalent to 10 h of the 50 h (one orbital period) subject to interference by sunlight. One of the three cavities cannot be used during these periods. Nonetheless, sunlight interference might be avoided by taking appropriate measures, such as changing the target attitudes of the satellites while one of the three cavities suspends scientific observations. 

It is important to note that an Earth eclipse can occur once per orbit. This is primarily because the orientation of the selected orbit is closely aligned with the ecliptic plane. As depicted in Fig. \ref{fig_d1} in Appendix \ref{Eclipse effects of the Earth}, the duration of these eclipses can extend up to approximately 100 minutes during both the vernal and autumnal equinoxes. Such eclipse events may limit the observation time owing to potential thermal instability in the satellite or constraints on the electrical power system. While this factor should be considered in the design of satellites, a detailed exploration of its implications is beyond the scope of this study.

\subsection{Astronomical interferometer}\label{Astronomical interferometer}
\begin{figure}
        \centering
        \includegraphics[width=1\columnwidth]{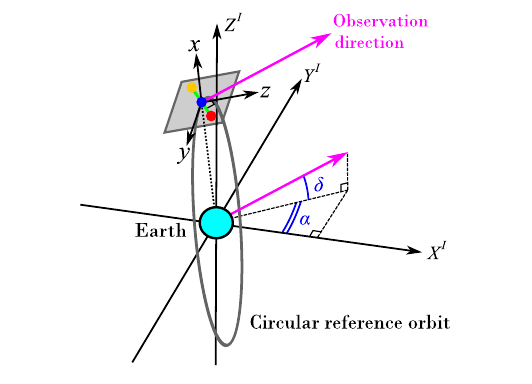}       
        \caption{Reference formation orbit for the astronomical interferometer. The two purple arrows are parallel and point in the same direction. }
        \label{fig7}
\end{figure}

The astronomical interferometer assumed in this study uses a linear formation under orthogonal inertial pointing, similar to nanosatellite astronomical interferometers \citep{hansen_ireland_2020, matsuo2022high}. The target observation direction in the ECI frame is given by $\boldsymbol{d}^I=[\cos{\delta}\cos{\alpha},\,\cos{\delta}\sin{\alpha},\,\sin{\delta}]^T$, where $\alpha$ and $\delta$ denote the right ascension and declination of the target observation direction, respectively. The reference formation can be configured by placing one spacecraft at the origin and the other two spacecraft symmetrically in the ($y-z$) plane in the LVLH frame \citep{hansen_ireland_2020}. A symmetrical spacecraft is placed by setting the parameters in Eq. (\ref{CW_bounded_solution}) as
$\rho_x=0$, $\rho_z = \rho_y\tan{p}$ ($-\pi/2<p<\pi/2$), and $\alpha_z=q$ $(0 \leq q < 2 \pi)$, and the other is placed symmetrically. Angles $p$ and $q$ are defined to satisfy the following relationships:
\begin{eqnarray}
\left\{
\begin{aligned}
\cos{p}&=\sin{\Delta\Omega_{\alpha}}\sin{I}\cos{\delta}+\cos{I}\sin{\delta}, \\
\sin{p}\sin{q}&=\sin{\Delta\Omega_{\alpha}}\cos{I}\cos{\delta} - \sin{I}\sin{\delta},   \\
\sin{p}\cos{q}&=\cos{\Delta\Omega_{\alpha}}\cos{\delta},
\end{aligned}\label{pq}
\right.
\end{eqnarray}
where $\Delta \Omega_{\alpha}=(\Omega - \alpha)$. The distances between spacecraft 1 (origin) and spacecrafts 2 and 3 are not necessarily constant; however, their optical path difference must be sufficiently small to attain the interference signals from the target directions. Although specific missions were not assumed in this study, the selected sizing parameter was $\rho_y=0.25$ km, similar to that in \citet{hansen_ireland_2020} as the typical mission. Figure \ref{fig7} illustrates the linear formation under orthogonal inertial pointing in Earth’s orbit. For this reference, the first and second time derivatives of $\rho_x$ and $\rho_y$ are zero, but $\dot{\rho}_z$, $\ddot{\rho}_z$, and $\dot{\omega}_t$ are nonzero; therefore, all four terms in Eq. (\ref{F_f2}) are nonzero. The first and second time derivatives of $\rho_z$ are given by $\dot{\rho}_z =(\dot{p}\rho_y/\cos^2{p})$ and $\ddot{\rho}_z=[(\ddot{p}+2\dot{p}^2\tan{p})\rho_y/\cos^2{p}]$, respectively, and $\dot{p}$ and $\ddot{p}$ can be calculated from Eq. (\ref{pq}). 

The most important observation condition for the orbital orientation of an astronomical interferometer is the coverage of the visible sky area. The key factors applied in this study are that the maximum angle between the observation target (star) and the Sun's directions is less than $\pi/3$ (condition 1) to maintain the solar arrays of each spacecraft facing to the Sun and keep the sunlight from interfering with the optical paths of the science instruments; and the clearance to the target star occultation by the Earth and its atmospheric thickness during one orbit (condition 2). These factors are derived from the same ideas as those in previous research \citep{hansen_ireland_2020}; however, condition 2 is a stricter rule because temporal occultation during one orbit is allowed in \citet{hansen_ireland_2020}, but not allowed in this study so that better observation conditions may be attained. In addition, this study imposes a further requirement that the angle ${|}p{|}$ be less than $\pi/3$ to regulate the formation size up to $2\rho_y$ (condition 3). The albedo effects of the Earth and Moon and the lunar occultation of the target stars were neglected in this study. Conditions 1--3 are given as follows:
\begin{eqnarray}
\left\{
\begin{aligned}
{\rm Condition 1:\,}&{0 \leq}\cos^{-1}{\left(-\boldsymbol{s}^I \cdot \boldsymbol{d}^I \right)}<\pi/3, \\
{\rm Condition 2:\,}& \cfrac{(R_e+h_t)}{a} < \cos{p}, \\
{\rm Condition 3:\,}&{|}p{|}<\pi/3,
\end{aligned}
\right.\label{ir_visibility_conditions}
\end{eqnarray}
where $\boldsymbol{s}^I$ is the unit Sun vector in the ECI frame, and $h_t$ is the Earth’s atmospheric thickness, assuming $h_t=100$ km. A spherical Earth is assumed to derive condition 2. Finally, the secular drift of $\Omega$ owing to Earth’s $J_2$ perturbation is actively used to increase the visible sky area during several years, so that the orbital inclination should be taken away from $\pm 90^{\circ}$ (see Eq. (\ref{J2_oe2}) for $\dot{\bar{\Omega}}$ in Appendix \ref{J2-perturbed orbital elements}). Considering these conditions, the orbital orientation parameters were selected as $I^{(s)}_0=70^{\circ}$ and $\Omega^{(s)}_0=0^{\circ}$. 

Lastly, the observation direction was set to $\alpha=19^{\rm h}\,20^{\rm m}$ and $\delta=20^{\circ}$ for the perturbing acceleration analysis. This selection was mainly aimed at understanding the typical acceleration magnitudes by taking the typical value of $p$ as approximately 20$^{\circ}$, and was not intended at particular observations of astronomical objects. 

\begin{figure*}
        \centering
        \resizebox{\hsize}{!}{\includegraphics[width=1\columnwidth]{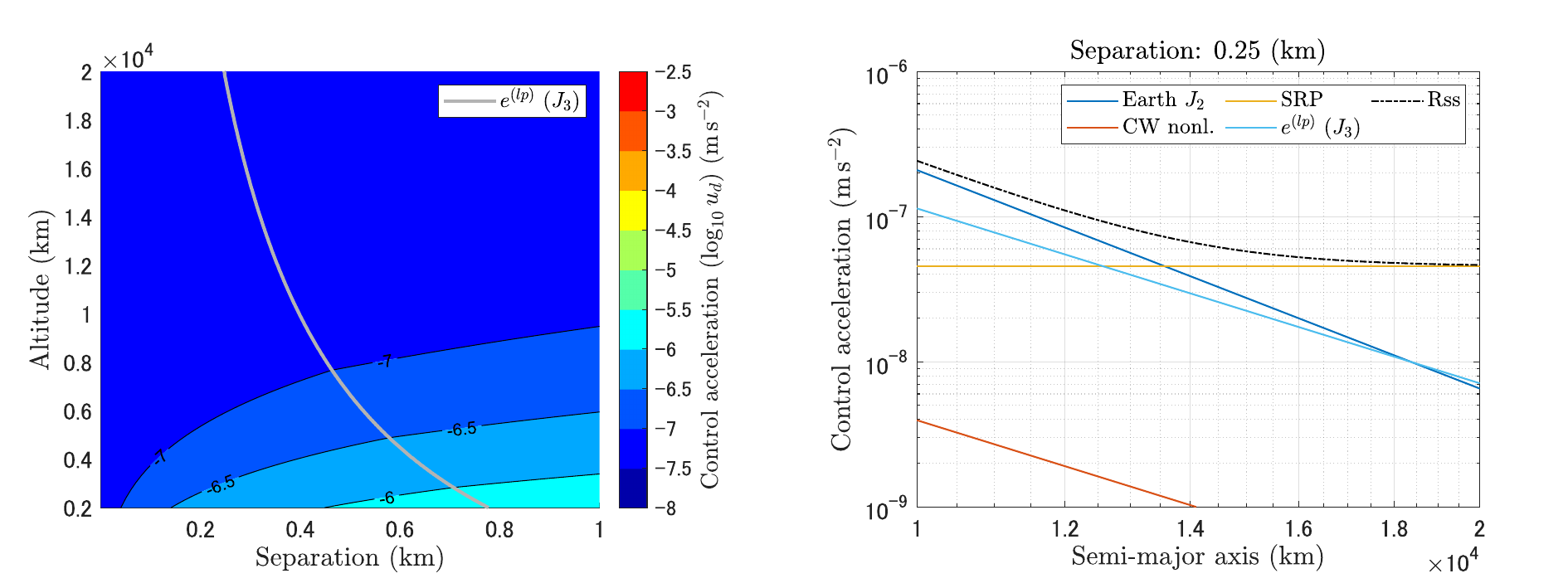}}        
        \caption{Control acceleration magnitudes for the linear astronomical interferometer. The left panel shows the contour map of the control acceleration magnitudes with respect to the separation (defined as $\rho_y$) and altitude. 
        For the long-period eccentricity drift owing to Earth’s $J_3$ gravity, the more propellant-saving approach appeared above the line (specified by $e^{(lp)}(J_3)$) when applying method (a) in Fig. \ref{fig1}, while it appeared below the line when applying method (b) in Fig. \ref{fig1}. 
        The borderlines for the secular and long-period drifts of the eccentricity vector owing to lunisolar gravity were abbreviated because their magnitudes were sufficiently smaller than $10^{-9}$ $\rm m\,s^{-2}$ in the plotted region. The right  panel shows the control acceleration magnitudes at the desired separation ($0.25$ km) with respect to the semimajor axis. The atmospheric drag, lunisolar gravity, and secular and long-period drifts of the eccentricity vector owing to the lunisolar gravity were removed from the figure because their magnitudes were sufficiently smaller than $10^{-9}$ $\rm m\,s^{-2}$. } \label{fig8}
\end{figure*}

The left panel of Fig. \ref{fig8} shows a contour map of the control acceleration magnitudes with respect to the separation and altitude of the astronomical interferometer. The right panel shows the control acceleration magnitude at the desired separation ($0.25$ km) with respect to the semimajor axis. A trend similar to that of B-DECIGO is shown in the left panel of Fig. \ref{fig8}: smaller control acceleration magnitudes appeared in the shorter-separation and higher-altitude region, and a small-disturbance region of less than $10^{-7}$ $\rm m\,s^{-2}$ was identified in medium Earth orbit. The control approach to the long-period drift of the eccentricity vector owing to Earth’s $J_3$ gravity was to compensate for the relative perturbing acceleration (in Fig. \ref{fig1} (b)) for the lower altitude and shorter separation and the absolute perturbing acceleration (in Fig. \ref{fig1} (a)) for the higher altitude and longer separation. 

By selecting the candidate Earth orbit for the astronomical interferometer at an altitude of 8,000 km ($a_0^{\rm (s)}=14,378$ km), the control acceleration magnitude was determined to be approximately $6\times 10^{-8}$ $\rm m\,s^{-2}$. This acceleration can be further broken down to compensate for the absolute and relative perturbations, each accounting for 31 \% ($||\boldsymbol{F}_c||$), 31 \% ($||\boldsymbol{f}_p||$), 26 \% ($||\boldsymbol{f}_o||$), and 12 \% ($||\boldsymbol{f}_f||$). 

\begin{figure}
        \centering
        \includegraphics[width=1\columnwidth]{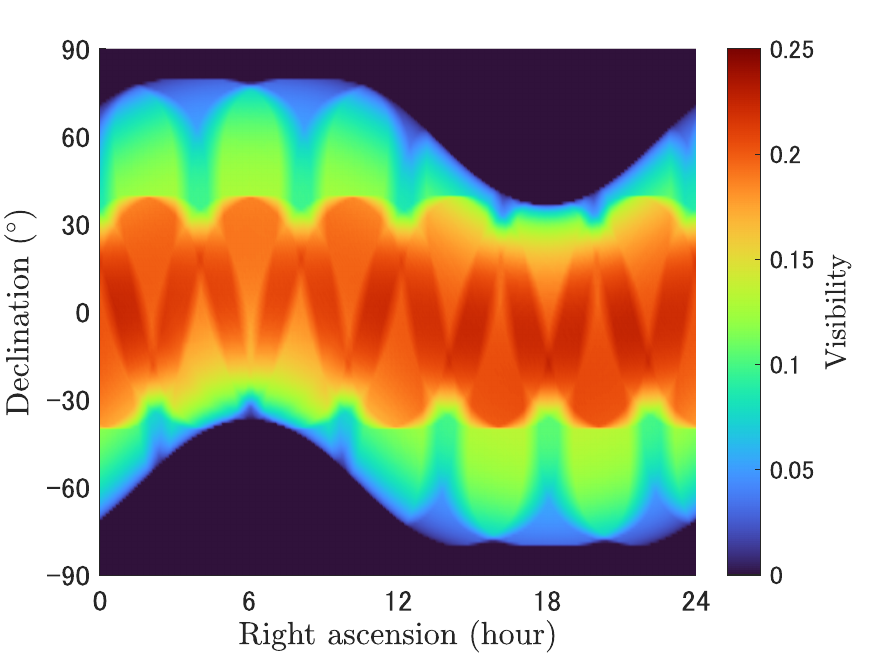} 
        \caption{Visible sky area for the astronomical interferometer at an altitude of 8,000 km and $I_0^{(s)}=70^\circ$. The color map indicates the ratio of the visibility in all sky directions over 5 years. }
        \label{fig9}
\end{figure}

Figure \ref{fig9} shows the visible sky area for an astronomical interferometer at an altitude of 8,000 km. The mission duration was assumed to be 5 years, where $\Omega$ drifted by $2\pi$ owing to the $J_2$ perturbation. The visible area ranged for any possible $\alpha$ within approximately $-75^{\circ}<{\delta}<75^{\circ}$ (depending on $\alpha$), and the maximum visibility was approximately $25$ \%. In particular, good visibility appeared within the area of $-30^{\circ}< {\delta} <30^{\circ}$. Although Earth eclipses do occur in the selected orbit for the astronomical interferometer, they are not a constant phenomenon in each orbit. The maximum duration of these eclipses is approximately 40 minutes, as illustrated in Fig. \ref{fig_d1} in Appendix \ref{Eclipse effects of the Earth}. 

\subsection{LEO environment}\label{LEO environment}
The use of LEOs has been considered in some formation-flying interferometry missions, particularly for astronomical interferometers \citep{hansen_ireland_2020, matsuo2022high}. However, their disturbance characteristics tended to be analyzed within mission-specific baseline lengths and orbits. Therefore, it remains unclear where and to what extent a small-disturbance environment can exist in LEO. Therefore, the LEO environment was assessed based on a perturbing acceleration analysis, assuming linear astronomical interferometers as potential applications. The Sun synchronous orbit was selected, where the RAAN drifts by $2\pi$ per year, and daylight is ensured in most seasons. The initial RAAN and inclination were set to $\Omega_{0}^{(s)}=10^{\circ}$ and $I_{0}^{(s)}=97.8^{\circ}$, respectively. The same observation direction in Sect. \ref{Astronomical interferometer} was used for the perturbing acceleration analysis. 

\begin{figure*}
        \centering
        \resizebox{\hsize}{!}{\includegraphics[width=1\columnwidth]{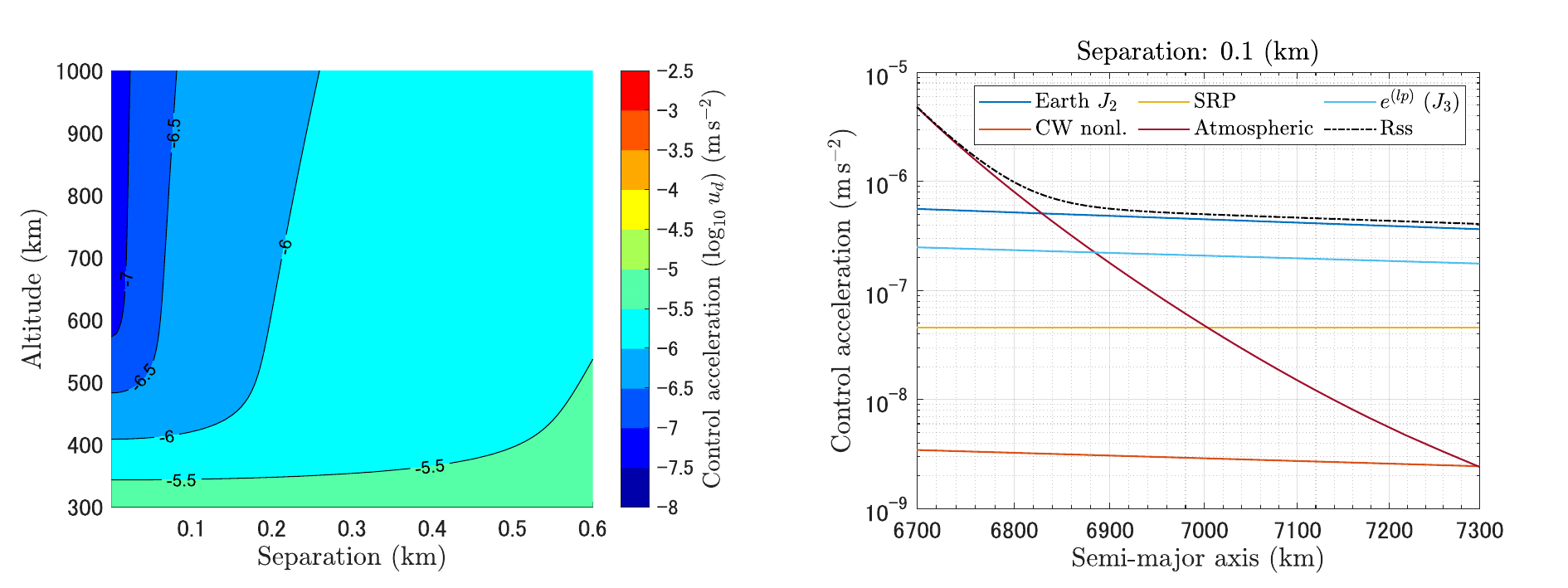}}        
    \caption{Control acceleration magnitudes for the linear astronomical interferometer in LEO. The left panel shows the contour map of the control acceleration magnitudes with respect to the separation (defined as $\rho_y$) and altitude. The control magnitudes against the secular and long-period drifts of the eccentricity vector owing to lunisolar gravity were sufficiently smaller than $10^{-9}$ $\rm m\,s^{-2}$ in the plotted region. The right panel shows the control acceleration magnitudes at the desired separation ($0.1$ km) with respect to the semimajor axis. The lunisolar gravity and the secular and long-period drifts of the eccentricity vector owing to lunisolar gravity were removed from the figure because their magnitudes were sufficiently smaller than $10^{-9}$ $\rm m\,s^{-2}$. } \label{fig10}
\end{figure*}

The left panel of Fig. \ref{fig10} shows a contour map of the control acceleration magnitudes with respect to separation (up to 1 km) and altitude (up to 1,000 km). The right panel shows the control acceleration magnitude at a specific separation ($0.1$ km) with respect to the semimajor axis. In contrast to the high and medium Earth orbits, the LEO has few small-disturbance regions of less than $10^{-7}$ $\rm m\,s^{-2}$. The largest perturbing acceleration sources were dominated by atmospheric drag (below approximately 600 km in altitude and 0.2 km in separation) and the Earth $J_2$ gravity potential for the other regions in Fig. \ref{fig10} left. However, a relatively small acceleration of $10^{-6.5}$ to $10^{-6}\,{\rm m\,s^{-2}}$ (considering $\Delta V$:10--30 $\rm m\,s^{-1}$ per year) can be attained in LEO at an altitude $>500$ km, with a separation of approximately $0.1$ km. For example, the control acceleration magnitude at an altitude of 600 km ($a_0^{\rm (s)}=6,978$ km) was approximately $5\times 10^{-7}$ $\rm m\,s^{-2}$. This acceleration can be further broken down to compensate for the absolute and relative perturbations, each accounting for 9 \% ($||\boldsymbol{F}_c||$), 52 \% ($||\boldsymbol{f}_p||$), 29 \% ($||\boldsymbol{f}_o||$), and 10 \% ($||\boldsymbol{f}_f||$). In the plotted region in Fig. \ref{fig10} left, the better mitigation approach to the $J_3$-perturbed long-period drift of the eccentricity vector was to compensate for the relative perturbing acceleration (in Fig. \ref{fig1} (b)). 

\begin{figure}
        \centering
        \includegraphics[width=1\columnwidth]{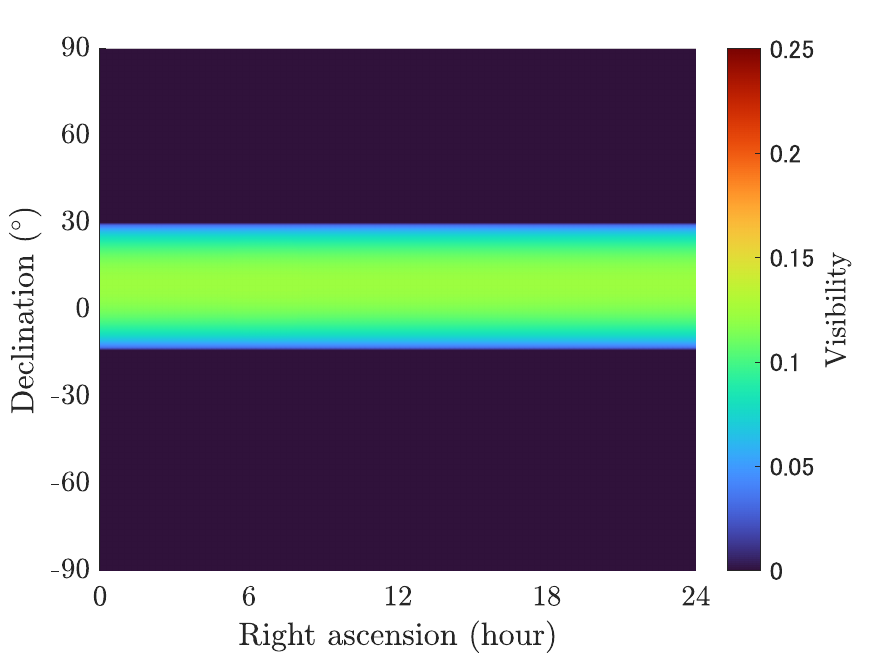} 
        \caption{Visible sky area for the astronomical interferometer at an altitude of 600 km and $I_0^{(s)}=97.8^\circ$. The color map indicates the ratio of the visibility in all sky directions over 1 year. }
        \label{fig11}
\end{figure}

Figure \ref{fig11} shows the visible sky area for an astronomical interferometer at an altitude of 600 km. The mission duration was assumed to be 1 year, where $\Omega$ drifted by $2\pi$ owing to the $J_2$ perturbation. In comparison to the results in Fig. \ref{fig9} and \citet{hansen_ireland_2020}, the visible area reduced within approximately $-14^{\circ}<{\delta}<30^{\circ}$ for any possible $\alpha$, and the maximum visibility decreased to approximately $12$ \%. This reduction in visibility is attributed to condition 2 in Eq. (\ref{ir_visibility_conditions}). When condition 2 is omitted, the visibility returns to levels similar to those found in Fig. \ref{fig9} and the work of \citet{hansen_ireland_2020}. Earth eclipses also occur, but they are seasonally limited. Their maximum duration is approximately 20 minutes, as detailed in Fig. \ref{fig_d1} in Appendix \ref{Eclipse effects of the Earth}. 

A negative aspect of using LEO for formation-flying interferometry is that the mission lifetime may be significantly shortened in comparison with those at higher geocentric orbits, owing to the limitation of the maximum load capability of the propellant. Another limitation is that precise formation control becomes more difficult with stronger perturbation effects. However, using LEO may outweigh these limitations from a benefit-to-cost perspective, particularly for small-class, experimental, and/or pathfinder missions. Moreover, using the latest low-thrust electrical propulsion with a high specific impulse can further improve the lifetime problem, as indicated in \citet{hansen_ireland_2020}. The last point to discuss is safe autonomous formation flying in proximity. In particular, proximity formation flying at 0.1 km or less is considered high-risk for collisions \citep{Monnier2019realistic}. In addition, the linear formation for the astronomical interferometer cannot apply the passively safe formation orbit known as the ``eccentricity/inclination vector separation'' technique \citep{damico2006proximity}. Therefore, active safety for collision-hazard detection and avoidance must be ensured using autonomous formation-flying functions. However, if this technical bottleneck is resolved, the LEO environment will become an ideal platform for demonstrating various technologies for future space interferometry.

\section{Conclusions}\label{Conclusions}
This study demonstrates the feasibility of formation-flying interferometry in geocentric orbits based on celestial mechanics. The perturbing accelerations to be mitigated by precise control can be categorized into four types: (1) absolute physical acceleration, (2) relative physical acceleration, (3) relative fictitious acceleration owing to the perturbed orbital motion of the chief, and (4) relative fictitious acceleration owing to the perturbed motion of the reference formation. In addition, the secular and/or long-period perturbations on the eccentricity vector owing to Earth $J_3$ and lunisolar gravity were unique because an efficient control approach could be changed to correct the absolute or relative perturbing accelerations. These results led to the characterization of the magnitude of each disturbance by the semimajor axis and the size of the formation. Consequently, small-perturbation conditions tended to appear in the higher-altitude and shorter-separation regions. Candidate orbits with a disturbance of less than $10^{-7}$ $\rm m\,s^{-2}$ were observed in the high Earth orbit for a laser-interferometric gravitational-wave telescope with a size of 100 km and in the medium Earth orbit for a linear astronomical interferometer with a size of 0.5 km. The LEO tended to have larger disturbance magnitudes; however, a LEO at a separation of approximately 0.1 km may be suitable for experimental purposes. These examples support the idea that the Earth’s orbit can be used for various types of formation-flying interferometry. 

\begin{acknowledgements}
      This study was supported by JSPS KAKENHI Grant Number JP23K13501. We would like to thank Dr. Shuichi Sato of Hosei University and Dr. Taro Matsuo of Nagoya University for their valuable advice on the formation control requirements for the space-based gravitational wave observatory and astronomical interferometer. 
\end{acknowledgements}

\bibliography{References}
\bibliographystyle{aa}

\begin{appendix}

\section{Nomenclature}\label{Nomenclature}

Table \ref{table6} lists the major symbols particularly used in the theoretical development of this study. 
\begin{table}[h!]
\caption{\label{table6} Nomenclature.}
\centering
\begin{tabular}{p{1.2cm}p{6.5cm}}
\hline \hline 
Symbol & Description \\ \hline
\multicolumn{2}{l}{Osculating orbital elements (OEs)} \\ 
$(\cdot)^{(s)}$ & Secular parts of osculating OEs, rad \\ 
$\delta(\cdot)^{(lp)}$ & Long-period parts of osculating OEs, rad\\
$\delta(\cdot)^{(sp)}$&Short-period parts of osculating OEs, rad\\
\multicolumn{2}{l}{Orbital parameter for the chief} \\ 
$a$&Semimajor axis, m \\
$\boldsymbol{e}$&Eccentricity vector ($\boldsymbol{e}=[e_x,\,e_y]^T$, $e=||\boldsymbol{e}||$) \\
$I$&Inclination, rad \\
$\Omega$&Right ascension of the ascending node, rad \\
$\omega$&Argument of perigee, rad \\
$\phi$&Mean argument of latitude, rad \\
$\theta$&Argument of latitude, rad \\
$n$& Mean motion, rad s$^{-1}$ \\
$n_l$&Long-period angular frequency,  rad s$^{-1}$ \\
$\boldsymbol{\omega}$&Orbital angular velocity vector, rad s$^{-1}$ \\
 &$(\boldsymbol{\omega}=[\omega_x,\,\omega_y,\,\omega_z]^T)$ \\
\multicolumn{2}{l}{Reference formation parameter} \\ 
$\Theta_x,\Theta_z$&Phase angles for the perturbed motion, rad\\
$\rho_x,\rho_y,\rho_z$&Size parameters, m\\
$\alpha_x,\alpha_z$&Phase parameters, rad\\
$\omega_t$&User-defined target angular velocity, rad s$^{-1}$\\
$\boldsymbol{r}_r$&Position vector of the deputy with respect to\\
&the chief, m $(\boldsymbol{r_r}=[x_r,\,y_r,\,z_r]^T)$ \\
\multicolumn{2}{l}{Physical and environmental parameter} \\ 
$\mu_b$&Third body's gravitational constant, m$^3$ s$^{-2}$\\
$\mu_e$&Earth gravitational constant, m$^3$ s$^{-2}$\\
$R_e$&Mean equatorial radius of Earth, m \\
$\rho$&Atmospheric density,  kg m$^{-3}$\\
$P_s$&Solar radiation pressure, Pa \\
\multicolumn{2}{l}{Spacecraft parameter} \\ 
$B_D$&Coefficient for drag, m$^2$ kg$^{-1}$ \\
$\delta B_D$&Differential value of $B_D$, m$^2$ kg$^{-1}$ \\
$B_s$&Coefficient for SRP, m$^2$ kg$^{-1}$ \\
$\delta B_s$&Differential value of $B_s$, m$^2$ kg$^{-1}$ \\
$C_s$&Radiation pressure coefficient \\
$C_d$&Drag coefficient \\
$S$&Cross-sectional area, m$^2$ \\
$m$&Spacecraft mass, kg \\
$\boldsymbol{s}$&Unit Sun vector \\
\multicolumn{2}{l}{Position} \\ 
$\boldsymbol{R}_b^I$&Position vector of the third body, m \\
$\boldsymbol{R}_c$&Position vector of the chief, m \\
$\boldsymbol{R}_d$&Position vector of the deputy, m \\
$\boldsymbol{r}$&Position vector of the deputy with respect to \\
&the chief, m $(\boldsymbol{r}=[x,\,y,\,z]^T)$ \\
\multicolumn{2}{l}{Acceleration} \\ 
$\boldsymbol{F}_c,\boldsymbol{f}_p,$&Perturbing acceleration vectors, m s$^{-2}$ \\
$\boldsymbol{f}_o,\boldsymbol{f}_f$& (see Table \ref{table1} for the definition.) \\
$\boldsymbol{u}_c$&Control acceleration vector applied to the chief, m s$^{-2}$ \\
$\boldsymbol{u}_d$&Control acceleration vector applied to the deputy, m s$^{-2}$ $(\boldsymbol{u}_d=[u_x,\,u_y,\,u_z]^T)$ \\
$\boldsymbol{u}_r$&Control acceleration vector to compensate for all relative perturbing accelerations, m s$^{-2}$ \\ \hline
\end{tabular}
\end{table}

\section{$J_2$-perturbed orbital elements}\label{J2-perturbed orbital elements}
For a mean circular orbit, the $J_2$-perturbed orbital motion of the chief up to the first order is given as follows:

\begin{eqnarray}
\left\{
\begin{aligned}
a&=a^{(s)}_0\left(1+\frac{3J}{2}\sin^2{I^{(s)}_0} \cos{2\phi^{(s)}}\right), \\
e_x&=\frac{J}{16} \left[14\sin^2{I^{(s)}_0}\cos{3\phi^{(s)}} + 3\left(3+5\cos{2I^{(s)}_0}\right)\cos{\phi^{(s)}} \right], \\
e_y&=\frac{J}{16} \left[14\sin^2{I^{(s)}_0}\sin{3\phi^{(s)}} + 3\left(1+7\cos{2I^{(s)}_0}\right)\sin{\phi^{(s)}} \right], \\
I&=I^{(s)}_0+\frac{3J}{8} \sin{2I^{(s)}_0} \cos{2\phi^{(s)}}, \\
\Omega&=\Omega^{(s)}_0+\dot{\bar{\Omega}}t+\frac{3J}{4} \cos{I^{(s)}_0} \sin{2\phi^{(s)}}, \\
\phi&=\phi^{(s)}_0+\dot{\bar{\phi}}t-\frac{3J}{8}\left(2-5\sin^2{I^{(s)}_0}\right) \sin{2\phi^{(s)}},
\end{aligned} \label{J2_oe}
\right.
\end{eqnarray}
where 
\begin{eqnarray}
\begin{aligned}
    J&=J_2\left(\cfrac{R_e}{a^{(s)}_0}\right)^2, \quad \dot{\bar{\phi}}=\bar{n}-\cfrac{3}{2}J\bar{n}\left(1-4\cos^2{I^{(s)}_0}\right), \\
    \dot{\bar{\Omega}}&=-\cfrac{3}{2}J \bar{n} \cos{I^{(s)}_0}, \quad \phi^{(s)}=\phi^{(s)}_0+\dot{\bar{\phi}}t, \quad \bar{n}=\sqrt{\cfrac{\mu}{{a^{(s)}}^3}}.
\end{aligned}\label{J2_oe2}
\end{eqnarray}

\section{Third-body-perturbed orbital elements}\label{Third-body-perturbed orbital elements}
The disturbing function owing to the third body can be written as follows \citep{Kozai1973}:
\begin{equation}
    {\mathcal R}=n_b^2 R_c^2 \beta \left(\cfrac{a_b}{R_b}\right)^3 \left[P_2(\cos{S_b})+\left(\cfrac{R_c}{R_b}\right) P_3(\cos{S_b})+ \cdots \right],
\end{equation}
where $a_b$, $R_b$, and $n_b$ are the semimajor axis, radius, and mean motion of the disturbing body, respectively; $P_2$, $P_3$, and $\cdots$ are Legendre polynomials; $\cos{S_b}$ is the direction cosine between the radius to the spacecraft and that to the disturbing body; $\beta=m_b/(m+m_b)$; and $m$ and $m_b$ are the masses of the Earth and disturbing body, respectively. The single-averaged disturbing function $\langle{\mathcal R}\rangle$ was derived by obtaining the average with respect to the mean anomaly $l$ of the chief. The double-averaged disturbing function $\langle\langle{\mathcal R}\rangle\rangle$ is derived by averaging $\langle{\mathcal R}\rangle$ with respect to the mean anomaly $l_b$ of the disturbing body. The long-period and short-period terms of the disturbing function are calculated using ${\mathcal R}^{(lp)}=(\langle{\mathcal R}\rangle-\langle\langle{\mathcal R}\rangle\rangle)$ and ${\mathcal R}^{(sp)}=({\mathcal R}-\langle{\mathcal R}\rangle)$, respectively. By replacing ${\mathcal R}$ with $\langle\langle{\mathcal R}\rangle\rangle$, ${\mathcal R}^{(lp)}$, and ${\mathcal R}^{(sp)}$ in Lagrange's equations, the orbital equations for the secular, long-period, and short-period motions are obtained, and the orbital elements can be calculated analytically by integrating the orbital equations with respect to time. 

If the terms with $e^2$ and higher are neglected, the single-averaged disturbing function up to $P_3(\cos{S_b})$ of the Legendre polynomials is written as follows \citep{Kozai1973}:
\begin{eqnarray}
\begin{aligned}
    \cfrac{\langle{\mathcal R}\rangle}{n_b^2a^2\beta}=&\cfrac{1}{4}\left(\cfrac{a_b}{R_b}\right)^3\left[3\left(A^2+B^2 \right)-2\right]
    -\cfrac{15}{16}e\left(\cfrac{a_b}{R_b}\right)^4\left(\cfrac{a}{a_b}\right) \times \\
    &\left[5\left(A^2+B^2\right)-4\right] \left(A\cos{\omega}+B\sin{\omega}\right).\label{Rls}
\end{aligned}
\end{eqnarray}
The $A$ and $B$ are defined as follows:
\begin{eqnarray}
\begin{aligned}
    A&=\cos{\delta_b}\cos{(\Omega - \alpha_b)},\\
    B&=\sin{\delta_b}\sin{I}-\cos{\delta_b}\cos{I}\sin{(\Omega - \alpha_b)},
\end{aligned}
\end{eqnarray}
where $\alpha_b$ and $\delta_b$ denote the right ascension and declination of the disturbing body in the ECI frame, respectively. Assuming that the eccentricity of the third body $(e_b)$ is small, the following approximations are applied:  
$(a_b/{R_b})^3 \approx (1+3e_b\cos{l_b})$ and $(a_b/{R_b})^4 \approx (1+4e_b\cos{l_b})$. This approximation is used to calculate $\langle\langle{\mathcal R}\rangle\rangle$. By contrast, a further approximation of $e_b=0$ (the circular-orbit assumption for the third body) is used to calculate ${\mathcal R}^{(lp)}$. The double-averaged disturbing function is calculated as follows:
\begin{eqnarray}
\begin{aligned}
    \cfrac{\langle\langle{\mathcal R}\rangle\rangle}{n_b^2a^2\beta}=&
    \cfrac{1}{8}\left[-4+3(k_1^2+k_2^2+k_3^2+k_4^2) \right]\\
    &-\cfrac{15}{64}e e_b \left(\cfrac{a}{a_b}\right) (K_1\cos{\omega}+K_2\sin{\omega}),
\end{aligned}
\end{eqnarray}
where 
\begin{eqnarray}
\begin{aligned}
k_1=& \cos{\Delta\Omega_b}\cos{\omega_b}+\sin{\Delta\Omega_b}\cos{I_b}\sin{\omega_b}, \\
k_2=&-\cos{I}\sin{\Delta\Omega_b}\cos{\omega_b}\\
    &+(\cos{\Delta\Omega_b}\cos{I_b}\cos{I}+\sin{I_b}\sin{I})\sin{\omega_b}, \\
k_3=&\cfrac{dk_1}{d\omega_b}, \quad
k_4=\cfrac{dk_2}{d\omega_b}, \quad \Delta \Omega_b=(\Omega-\Omega_b). 
\end{aligned}
\end{eqnarray}
The constants $K_1$ and $K_2$ are introduced to obtain compact expressions:
\begin{eqnarray}
\begin{aligned}
K_1&=-16k_1+5k_1\left[3(k_1^2+k_2^2+k_3^2)+k_4^2\right]+10k_2k_3k_4,\\
K_2&=-16k_2+5k_2\left[3(k_1^2+k_2^2+k_4^2)+k_3^2\right]+10k_1k_3k_4.
\end{aligned}
\end{eqnarray}

Subsequently, the orbital elements for the near-circular satellite motion are provided. The secular parts are
\begin{eqnarray}
\left\{
\begin{aligned}
    \dot{\bar{a}}&=0, \\
    \dot{\bar{e}}_{x}&=\cfrac{15}{64}e_b  \left(\cfrac{a}{a_b}\right) \Gamma_s K_2,\\    
    \dot{\bar{e}}_{y}&=-\cfrac{15}{64}e_b  \left(\cfrac{a}{a_b}\right)\Gamma_s K_1,\\    
    \dot{\bar{I}}&=-\cfrac{3}{4\sin{I}}\Gamma_s  \left(k_1k_5+k_2k_7+k_3k_6+k_4k_8\right), \\
    \dot{\bar{\Omega}}&=\cfrac{3}{4\sin{I}}\Gamma_s \left(k_2k_9+k_4k_{10}\right), \\ 
    \dot{\bar{\phi}}&=\bar{n}
    -\dot{\bar{\Omega}}\cos{I},\label{lunisolar_secular}
\end{aligned}
\right.
\end{eqnarray}
where
\begin{eqnarray}
\begin{aligned}
\Gamma_s=&\frac{n_b^2\beta}{n}, \quad
k_5=\cfrac{dk_1}{d\Omega}, \quad 
k_6=\cfrac{dk_5}{d\omega_b}, \quad 
k_7=\cfrac{dk_2}{d\Omega}, \\
k_8=&\cfrac{dk_7}{d\omega_b}, \quad
k_9=\cfrac{dk_2}{dI}, \quad
k_{10}=\cfrac{dk_9}{d\omega_b}.
\end{aligned}
\end{eqnarray}
The long-period parts are as follows:
\begin{eqnarray}
\left\{
\begin{aligned}
\delta a^{(lp)}=&0,  \\
\delta e^{(lp)}_{x}=&\cfrac{5}{64} \left(\cfrac{a}{a_b}\right) \Gamma^{(lp)}
\times\left\{
\begin{aligned}
    &5K_8\cos{3l_b}+5K_6\sin{3l_b} \\
    &-3K_4\cos{l_b}+3K_2\sin{l_b}
\end{aligned}
\right\},\\
\delta e^{(lp)}_{y}=&-\cfrac{5}{64}\left(\cfrac{a}{a_b}\right)\Gamma^{(lp)}
\times\left\{
\begin{aligned}
    &5K_7\cos{3l_b}+5K_5\sin{3l_b} \\
    &-3K_3\cos{l_b}+3K_1\sin{l_b}
\end{aligned}
\right\},\\
\delta I^{(lp)}=&\cfrac{3}{8\sin{I}}\Gamma^{(lp)}
\times\\
&\left\{
\begin{aligned}
    &(k_1k_6+k_2k_8+k_3k_5+k_4k_7)\cos{2l_b} \\
    &+(-k_1k_5-k_2k_7+k_3k_6+k_4k_8)\sin{2l_b}
\end{aligned}
\right\},\\
\delta \Omega^{(lp)}=&\cfrac{3}{8\sin{I}}\Gamma^{(lp)}
\times\left\{
\begin{aligned}
    &-(k_2k_{10}+k_4k_9)\cos{2l_b} \\
    &+(k_2k_9-k_4k_{10})\sin{2l_b}
\end{aligned}
\right\},\\
\delta \phi^{(lp)}=&\cfrac{3}{4} \Gamma^{(lp)} 
\times \\
&\left\{
\begin{aligned}
    &(k_1k_3+k_2k_4)\cos{2l_b} \\
    &-2(k_1^2+k_2^2-k_3^2-k_4^2)\sin{2l_b}
\end{aligned}
\right\} -\delta \Omega^{(lp)} \cos{I},
\end{aligned}
\right.
\end{eqnarray}
where 
\begin{eqnarray}
\begin{aligned}
    \Gamma^{(lp)}&=\frac{n_b\beta}{n},\\    K_3&=-16k_3+5k_3[3(k_1^2+k_3^2+k_4^2)+k_2^2]+10k_1k_2k_4,\\
    K_4&=-16k_4+5k_4[3(k_2^2+k_3^2+k_4^2)+k_1^2]+10k_1k_2k_3,\\
    K_5&=k_1(k_1^2+k_2^2-3k_3^2-k_4^2)-2k_2k_3k_4,\\
    K_6&=k_2(k_1^2+k_2^2-k_3^2-3k_4^2)-2k_1k_3k_4,\\
    K_7&=k_3(-3k_1^2-k_2^2+k_3^2+k_4^2)-2k_1k_2k_4,\\
    K_8&=k_4(-k_1^2-3k_2^2+k_3^2+k_4^2)-2k_1k_2k_3.
\end{aligned}
\end{eqnarray}

The short-period parts are as follows:
\begin{eqnarray}
    \delta (\cdot)^{(sp)}&=&\delta (\cdot)^{(sp2)}+(\cdot)^{(sp3)},
\end{eqnarray}
and they are given by
\begin{eqnarray}
\left\{
\begin{aligned}
\delta a^{(sp2)}=& \cfrac{3}{2}a\Gamma^{(sp2)}
\left[(A^2-B^2)\cos{2\phi}+2AB\sin{2\phi} \right], \\
\delta e^{(sp2)}_x=&\cfrac{1}{4}\Gamma^{(sp2)}
\times\\
&\left\{
\begin{aligned}
    &(A^2-B^2)\cos{3\phi}+2AB\sin{3\phi} \\
    &+\left[4+3(A^2-5B^2)\right]\cos{\phi}+18AB\sin{\phi} 
\end{aligned}
\right\},\\
\delta e^{(sp2)}_y=&\cfrac{1}{4}\Gamma^{(sp2)}
\times\\
&\left\{
\begin{aligned}
    &-2AB\cos{3\phi}+(A^2-B^2)\sin{3\phi}  \\
    &+18AB\cos{\phi}+\left[4+3(-5A^2+B^2)\right]\sin{\phi} 
\end{aligned}
\right\},\\
\delta I^{(sp2)}=&\cfrac{3}{4\sin{I}}\Gamma^{(sp2)}
\times\\
&\left\{
\begin{aligned}
    &\left[Ak_{12}+Bk_{11}+(A^2-B^2)\cos{I}\right]\cos{2\phi} \\
    &+\left(-Ak_{11}+Bk_{12}+2AB\cos{I}\right)\sin{2\phi}
\end{aligned}
\right\},\\
\delta \Omega^{(sp2)}=&-\cfrac{3}{4\sin{I}}\Gamma^{(sp2)}
k_{13}\left(A\cos{2\phi}+B\sin{2\phi} \right), \\
\delta \phi^{(sp2)}=& \cfrac{21}{8}\Gamma^{(sp2)}
\left[2AB\cos{2\phi}+(-A^2+B^2)\sin{2\phi} \right]\\
&-\delta \Omega^{(sp2)}\cos{I},
\end{aligned}
\right.
\end{eqnarray}

\clearpage
and
\begin{eqnarray}
\left\{
\begin{aligned}
\delta a^{(sp3)}=&\cfrac{1}{4}a\Gamma^{(sp3)}
\times\left\{
\begin{aligned}
    &5A(A^2-3B^2)\cos{3\phi}\\
    &+5B(3A^2-B^2)\sin{3\phi} \\
    &+3A\left[-4+5(A^2+B^2)\right]\cos{\phi}\\
    &+3B\left[-4+5(A^2+B^2)\right]\sin{\phi} 
\end{aligned}
\right\},\\
\delta e^{(sp3)}_x=&\cfrac{3}{64}\Gamma^{(sp3)}
\times\left\{
\begin{aligned}
    &5A(A^2-3B^2)\cos{4\phi}\\
    &+5B(3A^2-B^2)\sin{4\phi} \\
    &+4A\left[2+5(A^2-5B^2)\right]\cos{2\phi}\\
    &+8B\left[1+5(2A^2-B^2)\right]\sin{2\phi} 
\end{aligned}
\right\},\\
\delta e^{(sp3)}_y=&\cfrac{3}{64}\Gamma^{(sp3)}
\times\left\{
\begin{aligned}
    &5B(-3A^2+B^2)\cos{4\phi}\\
    &+5A(A^2-3B^2)\sin{4\phi} \\
    &-4B\left[2-5(5A^2-B^2)\right]\cos{2\phi}\\
    &{+}8A\left[1-5(A^2-2B^2)\right]\sin{2\phi} 
\end{aligned}
\right\},\\
\delta I^{(sp3)}=&\cfrac{1}{8\sin{I}}\Gamma^{(sp3)}
\times \left\{
\begin{aligned}
    &5K_{9}\cos{3\phi}+5K_{10}\sin{3\phi} \\
    &+3K_{11}\cos{\phi}-3K_{12}\sin{\phi} 
\end{aligned}
\right\},\\
\delta \Omega^{(sp3)}=&-\cfrac{1}{8\sin{I}}\Gamma^{(sp3)}
\times \\
&k_{13}\left\{
\begin{aligned}
    &5(A^2-B^2)\cos{3\phi}\\
    &+10AB\sin{3\phi} \\
    &+3\left[-4+5(A^2+3B^2)\right]\cos{\phi}\\
    &-30AB\sin{\phi} 
\end{aligned}
\right\},\\
\delta \phi^{(sp3)}=&\cfrac{3}{8}\Gamma^{(sp3)}
\times \left\{
\begin{aligned}
    &5B(3A^2-B^2)\cos{3\phi}\\
    &-5A(A^2-3B^2)\sin{3\phi} \\
    &+9B\left[-4+5(A^2+B^2)\right]\cos{\phi}\\
    &-9A\left[-4+5(A^2+B^2)\right]\sin{\phi}
\end{aligned}
\right\}\\
&-\delta \Omega^{(sp3)}\cos{I},
\end{aligned}
\right.
\end{eqnarray}
where
\begin{eqnarray}
\begin{aligned}
    \Gamma^{(sp2)}=&\left(\cfrac{n_b}{n}\right)^2\left(\cfrac{a_b}{R_b}\right)^3{\beta}, \quad
    \Gamma^{(sp3)}=\left(\cfrac{a}{R_b}\right)\Gamma^{(sp2)}, \\
    k_{11}=&\cfrac{\partial A}{\partial \Omega}, \quad
    k_{12}=\cfrac{\partial B}{\partial \Omega}, \quad
    k_{13}=\cfrac{\partial B}{\partial I}, \\
    K_9=&2ABk_{11}+(A^2-B^2)k_{12}+A(A^2-3B^2)\cos{I},\\
    K_{10}=&2ABk_{12}-(A^2-B^2)k_{11}+B(3A^2-B^2)\cos{I},\\
    K_{11}=&10ABk_{11}+\left[-4+5(A^2+3B^2)\right]k_{12}\\
    &-A\left[4-5(A^2+B^2)\right]\cos{I},\\
    K_{12}=&10ABk_{12}+\left[-4+5(3A^2+B^2)\right]k_{11}\\
    &+B\left[4-5(A^2+B^2)\right]\cos{I}. 
\end{aligned}
\end{eqnarray}

To calculate the osculating orbital elements, the secular orbital elements are first computed. Then, the long-period variation $\delta (\cdot)^{(lp)}$ is calculated using the secular orbital elements, and the short-period variation $\delta (\cdot)^{(sp)}$ is calculated using the secular and long-period orbital elements. 

\section{Eclipse effects of the Earth}\label{Eclipse effects of the Earth}
\begin{figure}
        \centering
        \includegraphics[width=1\columnwidth]{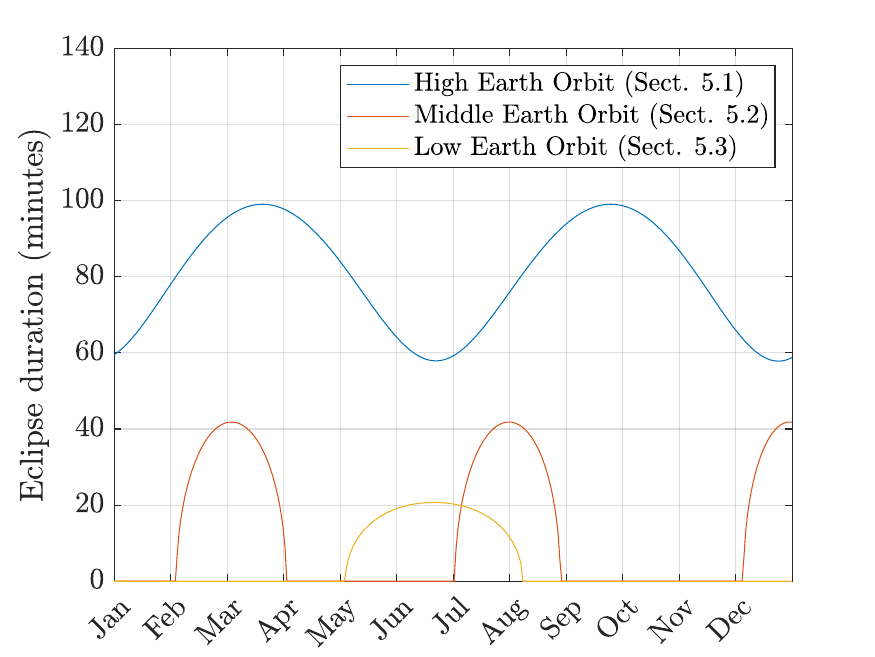} 
        \caption{Eclipse effects of the Earth in various selected orbits. } \label{fig_d1}
\end{figure}

Figure \ref{fig_d1} illustrates the Earth's eclipse effects spanning from January 1, 2024, 00:00:00 UT to December 31, 2024, 24:00:00 UT for the orbits selected in Sect. \ref{Results and discussion}. The three lines in Fig. \ref{fig_d1} represent the annual duration of eclipses per orbit for the orbits discussed in Sects. \ref{Gravitational-wave telescope} (depicted in blue), \ref{Astronomical interferometer} (in orange), and \ref{LEO environment} (in yellow). These durations were computed under the assumption of circular orbits. For low and medium Earth orbits, $(J_2$)-perturbed secular motion was considered, whereas high Earth orbits were assumed to exhibit unperturbed motion. Furthermore, only the effects of the umbra were taken into account.

\end{appendix}
\end{document}